\documentclass[aps,prd,showpacs,nofootinbib,twocolumn]{revtex4}
\usepackage{latexsym}
\usepackage{amsmath,amsfonts}
\usepackage{amsbsy}
\usepackage{mathrsfs}
\usepackage{color}
\usepackage{float}
\usepackage{dsfont}
\usepackage{psfrag}

\usepackage{enumerate}

\usepackage{amsmath,amssymb,calc,amsfonts}
\usepackage{latexsym}
\usepackage{hyperref}

\usepackage{graphicx,calc,epsfig}

\begin{document}

\title{Expressions for optical scalars and deflection angle at second order in terms of curvature scalars

}

\date{\today}

\author{Gabriel Crisnejo$^1$ and Emanuel Gallo$^1,^2$}

\affiliation{$^1$FaMAF, UNC; Ciudad Universitaria, (5000) C\'ordoba, Argentina.\\	
 $^2$Instituto de F\'isica Enrique Gaviola (IFEG), CONICET, \\
Ciudad Universitaria, (5000) C\'ordoba, Argentina. }

\begin{abstract}
 We present formal expressions for the optical scalars in terms of the curvature scalars  in the weak gravitational lensing regime at second order in perturbations of a flat background without mentioning the extension of the lens or their shape. Also, by considering the thin lens approximation for static and axially symmetric configurations we obtain an expression 
for the second-order deflection angle which generalizes our previous result presented in \cite{PhysRevD.83.083007}. As applications of these formulas we compute the optical scalars for some known family of metrics and we recover expressions for the deflection angle. In contrast to other works in the subject, our formalism allows a straightforward identification of how the different components of the curvature tensor contribute to the optical scalars and deflection angle. We also discuss in what sense the Schwarzschild solution can be thought as a true thin lens at second order.
\end{abstract}


\maketitle


\section{Introduction}\label{section1}
The phenomenon of gravitational lensing has become an active area of research at least from the beginning of the 1980's when the observation of the first astrophysical gravitational lens was announced\cite{1979Natur.279..381W}. Since then, a plethora of different astrophysical and cosmological gravitational lenses have been discovered and studied in depth. For these reasons, even today, this subject is still being theoretical considered due to the numerous applications in both astrophysical and cosmological settings,
such as the study of CMB power spectrum\cite{Lewis:2006fu,Nguyen:2017zqu,Marozzi:2016und,Peloton:2016kbw,Fabbian:2017wfp,Pratten:2016dsm}, the distance-redshift relationship\cite{Kaiser:2015iia,Fleury:2016fda}, or the estimation of mass content of groups or clusters of galaxies\cite{Hoekstra:2013via,2015IAUS..311...86M,Giocoli:2013tga}, to cite some examples.

In \cite{PhysRevD.83.083007}, we presented a new approach to the study of gravitational lensing in the weak field regime, which proved to be very useful since gives explicit gauge invariant expressions 
not only for the optical scalars but also to the deflection angle. These expressions can be easily implemented in any gauge.  In particular, in that reference we also presented expressions for these optical quantities in terms of the components of the energy-momentum tensor instead of the usual presentation in terms of metric potentials.
Applications of the formalism can be found in \cite{Bozza:2015haa, Gallo:2011hi}, and its results have been recently extended to the cosmological 
framework\cite{Boero:2016nrd}.

In the case of an axially symmetric lens configuration, it was also shown that at first order the deflection angle $\alpha$ in a weak field regime of an asymptotically flat spacetime can be written in terms of the impact parameter $J$,  and the projected Ricci and Weyl scalars $\hat\Phi_{00}$ and $\hat\Psi_0=-\hat\psi_0e^{2i\theta}$ (We refer to \cite{PhysRevD.83.083007} for more details) in a very compact form which we reproduce here
\begin{equation}\label{angulito}
\alpha(J)=J\left(\hat\Phi_{00}(J)+\hat\psi_0(J)\right).
\end{equation}

On the other hand, the second-order gravitational lens theory, that is, the study of the optical scalars at second order in perturbation of a given background metric and its applications has been extensively covered.  For example, in \cite{Vanderveld:2011sj} the weak field lens equation at second order in $\tilde\varepsilon=G/c^{2}$ has been worked out in the post-Newtonian formalism. Fritelli {\it et al.} \cite{Frittelli:1999yf} studied the exact lens equation in Schwarzschild spacetime and, in particular, the lens equation at second order in $\tilde\varepsilon$. The light deflection at second order in $\tilde\varepsilon$ has been studied in \cite{Brugmann:2005ft} for a system of two bounded point masses. In the cosmological context we found several works concerning the second order effect of weak lensing in the CMB power spectrum\cite{Hagstotz:2014qea, 1475-7516-2015-06-050, Kaiser:2015iia, Marozzi:2016uob, Petri:2016qya, Schaefer:2005up, Cooray:2002mj,Marozzi:2016qxl}. A general expression for  the  convergence  at  second  order in the Poisson gauge can
be  obtained  using  the  area  distance~\cite{BenDayan:2012wi,Fanizza:2013doa,Marozzi:2014kua} or alternatively using the so-called galaxy number counts~\cite{DiDio:2014lka,DiDio:2015bua}. In the same gauge a general expression for the shear at second order can be found in~\cite{Bernardeau-2010}.

However, in all these works the different expressions for the optical scalars or the deflection angle  are generally written in a particular gauge or a given family of gauges. 
Although all these studies are very useful and relevant in their respective range of validity, it is our intention to extend some of  the results presented in \cite{PhysRevD.83.083007}. In particular, we will obtain expressions for the optical scalars in terms of curvature scalars which take into account second order perturbations of a flat metric, and which are independent of the extending or shape of the lens. It is also our intention to generalize \eqref{angulito} at second order. . 

One of the advantages of the present approach is that allows us to recognize how the different aspects of the curvature of the spacetime contribute to the deflection angle and optical scalars and on the other hand they are ready for use in any gauge.

The organization of the paper is as follows. In Sec.~\ref{section2} we present a short review of the main equations of weak gravitational lensing, including a brief discussion of the geodesic deviation equation and we solve the geodesic deviation equation 
up to second-order in perturbation of a flat metric and we obtain, via the amplification matrix, general expressions for the optical scalars.  In Sec.~\ref{section4}, we focus on the thin lens approximation and give an explicit expression for the deflection angle for static and axially symmetric lenses which generalizes \eqref{angulito}.
In Sec.~\ref{Applications}, we provide some examples, applying the formalism to the description of the optical scalars for some known metrics and recovering expressions for the weak lensing quantities in those spacetimes. We conclude with general remarks and comments in the last section. Some auxiliary but relevant relations have been included
in the four Appendixes.

\section{The geodesic deviation equation: second order solution}\label{section2}
\subsection{Preliminaries}
As is well-known, the weak lens equation relates the angular position $\beta^a$ of the source (which should be the observed angular position if there were no gravitational lens) and the actual observed angular position $\theta^a$ due to the presence of the lens, 
\begin{equation}\label{eq: 0.1}
\beta^{a} = \theta^{a} - \frac{\lambda_{ls}}{\lambda_{s}} \alpha^{a};
\end{equation}
with $\alpha$ the deflection angle~\cite{Ehlers:1992dau}.
The differential of the Eq.\eqref{eq: 0.1} can be written as:
\begin{equation}\label{eq:0.00}
\delta \beta^{a} = A^{a}{}_{b} \ \delta \theta^{b},
\end{equation}
where $A^{a}{}_{b}$ is called the \emph{amplification matrix} and is given by:
\begin{equation}\label{Aij}
A^a{}_b=\begin{pmatrix}
1 - \kappa - \gamma_1   &   -\gamma_2 - \hat{\omega}   \\
-\gamma_2 + \hat{\omega}   &   1 - \kappa + \gamma_1   \\
\end{pmatrix};
\end{equation}
where $\kappa$, $\gamma \equiv \gamma_1 + i \, \gamma_2$ and $\hat{\omega}$ are called \emph{optical scalars}: convergence, shear and rotation, respectively.

A powerful way to study weak gravitational lensing is through the use of the geodesic deviation equation. A complete discussion of this equation and its use in the description of gravitational lensing can be found in~\cite{Bartelmann:2010fz,Seitz:1994xf}.

Let us consider a null geodesic congruence starting at the source position  $S$ and ending at the observer position $O$. That is, a null congruence belonging to the past null cone of $O$.
The tangent vector to a fiducial null geodesic of this congurence is given by  $\ell = \frac{\partial}{\partial \lambda}$.
At the position of the observer, we can construct a null tetrad $\{\ell^a,n^a,m^a,\bar{m}^a\}$ satisfying the standard normalization conditions, and with $m^a, \bar{m}^{a}$ complex null vectors which are orthogonal to the four-velocity $u^a$ of $O$.
The deviation vector connecting two neighboring null geodesics in the congruence can be expressed by
\begin{equation}\label{zeta_forma}
 \zeta^{a} = \zeta \bar{m}^{a} + \bar{\zeta} m^{a} + \zeta_{\ell} \, \ell^{a}.
\end{equation}

As it is well discussed in the literature, the geodesic deviation equation can be written as~\cite{Bartelmann:2010fz,Seitz:1994xf, Frittelli:2000bc}:
\begin{equation}\label{dg:so}
  \ell(\ell(\mathcal{X})) = - Q \mathcal{X};
\end{equation}
where
\begin{equation}
\mathcal{X} =
\begin{pmatrix}
\zeta \\
\bar{\zeta}
\end{pmatrix},
\end{equation}
and 
\begin{equation}		
Q = \begin{pmatrix}
\Phi_{00} & \Psi_{0} \\
\bar{\Psi}_{0} & \Phi_{00} 
\end{pmatrix},
\end{equation}
with
\begin{equation}
\Phi_{00} = -\frac{1}{2} R_{ab} \ell^{a} \ell^{b} , \ \ \ \Psi_{0} = C_{abcd} \ell^{a} m^{b} \ell^{c} m^{d}.
\end{equation}

\subsection{Second-order solution}\label{subsection3}

Now we will solve the Eq.\eqref{dg:so} at second order in a perturbation of a flat spacetime following the iterative method used in \cite{PhysRevD.83.083007}.

We define
\begin{equation}
X =
\begin{pmatrix}
\mathcal{X} \\
\mathcal{V}
\end{pmatrix}, \ \ \ 
\mathcal{V} = \frac{d \mathcal{X}}{d \lambda}.
\end{equation}
Therefore,  Eq.(\ref{dg:so}) is reduced to a first order differential equation given by:
\begin{equation}\label{dg:fo}
\ell(X) = \mathbb{A} \, X,
\end{equation}
where
\begin{equation}
\mathbb{A} =
\begin{pmatrix}
0 & \mathds{1} \\
-Q & 0 
\end{pmatrix}.
\end{equation} 
Following \cite{PhysRevD.83.083007} we start from a seed $X_{0}$, and we will perform the integrations from the observer position $\lambda_{o}$ (which can be taken without a loss of generality to be zero ) to the source position in $\lambda_{s}$. 
The beam has initially a vanishing departure, and therefore the seed is taken as (see ~\cite{PhysRevD.83.083007} for details)
\begin{equation}
X_{0} = \left(
\begin{array}{c}
0 \\
\mathcal{V}(0)
\end{array}\right).
\end{equation}  
Then we construct the following sequence of approximate solutions:
\begin{equation}\label{eq:2.1}	  
	X_{1}(\lambda_s) = X_{0} + \int^{\lambda_s}_{0} \mathbb{A}(\lambda{'}) X_{0} d\lambda{'}, 
\end{equation}
\begin{equation}\label{eq:2.2}
	X_{2}(\lambda_s) = X_{0} + \int^{\lambda_s}_{0} \mathbb{A}(\lambda{'}) X_{1} d\lambda{'},
\end{equation}
and so on. By replacing \eqref{eq:2.1} into \eqref{eq:2.2} we get
\begin{equation}
	X_{2}(\lambda_s) = X_{1} + \int^{\lambda_s}_{0}\int^{\lambda{'}}_{0} \mathbb{A}(\lambda{'})\mathbb{A}(\lambda{''}) X_{0} d\lambda{''} d\lambda{'};
\end{equation}
in this way we can obtain a solution of \eqref{dg:fo} to the desired order. We see that in the $X_{\emph{n}}$ step there will be \emph{n} products of matrices $\mathbb{A}$. As we are interested in second-order solutions, 
we only need to compute up to $X_{5}$ because in the next step only appear cubic quantities in Q.
Accordingly, by considering up to quadratic terms in Q we get
\begin{equation}\label{eq:2.3}
\begin{aligned}
&\mathcal{X}(\lambda_s) =  \bigg[ \mathds{1}\lambda_{s} - \int^{\lambda_s}_{0} \int^{\lambda{'}}_{0}\lambda{''} Q(\lambda{''}) 
		    d\lambda{''} d\lambda{'} \\
		    &+ \int_{0}^{\lambda_s}\int_{0}^{\lambda{'}}\int_{0}^{\lambda{''}}\int_{\lambda_ {0}}^{\lambda{'''}}\lambda{''''}
		      Q(\lambda{''}) Q(\lambda{''''}) d\lambda{''''} d\lambda{'''}d\lambda{''}d\lambda{'} \bigg]  \\
		      &\times \mathcal{V}(0),
\end{aligned}							
\end{equation}
On the other hand, if the metric were flat $(Q = 0)$ then we should have
\begin{equation}
\mathcal{X}_{s} \equiv	\mathcal{X}(\lambda_{s}) = \lambda_{s} \mathcal{V}(0) \ \ \Rightarrow \ \ \mathcal{V}(0) 
= \frac{\mathcal{X}(\lambda_{s})}{\lambda_{s}}.
\end{equation}
But in the presence of a gravitational lens, if an observer sees an image of size $\mathcal{X}_{o}$, which means 
$\mathcal{X}_{o} \equiv\lambda_{s} \mathcal{V}_{0}$ then it should be produced by a source of size 
$\mathcal{X}_{s} = \mathcal{X}(\lambda_{s})$, as described by Eq.\eqref{eq:2.3}. 

Using the following two relations obtained by integration by parts [see Appendix \eqref{app:int}]:
\begin{equation}\label{identity1}
\int_{0}^{\lambda_{s}}\int_{0}^{\lambda{'}}\lambda{''}  Q(\lambda{''}) d\lambda{''}d\lambda{'} 
= 	\int^{\lambda_{s}}_{0} \lambda(\lambda_{s} - \lambda) Q(\lambda) d\lambda,
\end{equation}
\begin{equation}\label{identity2}
\begin{aligned}
&\int^{\lambda_{s}}_{0} \int^{\lambda{'}}_{0} \int^{\lambda{''}}_{0} \lambda{'''} (\lambda{''}-\lambda{'''})
	Q(\lambda{''}) Q(\lambda{'''}) d\lambda{'''} d\lambda{''} d\lambda{'} \\
	&= \int_{0}^{\lambda_{s}}\int_{0}^{\lambda}\lambda{'}(\lambda_{s}-\lambda)(\lambda-\lambda{'})Q(\lambda)Q(\lambda{'})d\lambda{'}d\lambda,
\end{aligned}
\end{equation}
we finally obtain
\begin{equation}\label{eq:2.44}
\begin{aligned}
	\mathcal{X}_{s} &=  \bigg[\mathds{1} - \frac{1}{\lambda_{s}} \int^{\lambda_{s}}_{0} \lambda (\lambda_{s} - \lambda) Q(\lambda) d\lambda \\
	&+ \frac{1}{\lambda_{s}} \int_{0}^{\lambda_{s}}\int_{0}^{\lambda}\lambda{'}(\lambda_{s}-\lambda)(\lambda-\lambda{'})Q(\lambda)Q(\lambda{'})d\lambda{'}d\lambda \bigg] \mathcal{X}_{o};
\end{aligned}
\end{equation}
where the deviation vector at the source position $\mathcal{X}_{s}$ and at the observer position $\mathcal{X}_{0}$ are given by
\begin{equation}\label{eqxs}
\mathcal{X}_{s}=\left( 
\begin{array}{c}
\zeta_{s} \\
\bar{\zeta}_{s}	
\end{array}\right), \ \ \ \ \ \ \ \ \ \ \mathcal{X}_{0}=\left( 
\begin{array}{c}
\zeta_{o} \\
\bar{\zeta}_{o}	
\end{array}\right).
\end{equation}  
  
By replacing \eqref{eqxs} into \eqref{eq:2.44} and using the explicit expression for the $Q$ matrix we have:
\begin{equation}\label{eq:2.41}
\begin{aligned}
\zeta_{s}  =&  \bigg[ 1 - \frac{1}{\lambda_{s}} \int^{\lambda_{s}}_{0} \lambda (\lambda_{s} - \lambda) \Phi_{00}(\lambda) d\lambda 
	   +  \frac{1}{\lambda_{s}} \int^{\lambda_{s}}_{0} \int^{\lambda}_{0} \lambda{'} \\
	   &(\lambda_{s} - \lambda)(\lambda-\lambda{'}) 
           \bigg(\Phi_{00}(\lambda) \Phi_{00}(\lambda{'})  
	   +  \Psi_{0}(\lambda)\bar{\Psi}_{0} (\lambda{'})\bigg) \\ 
	   &d\lambda{'}d\lambda \bigg] \zeta_{o} 
	   +  \bigg[- \frac{1}{\lambda_{s}} \int_{0}^{\lambda_{s}} \lambda 
	   (\lambda_{s} - \lambda) \Psi_{0}(\lambda) d\lambda  \\
	   &+ \frac{1}{\lambda_{s}} \int^{\lambda_{s}}_{0} \int^{\lambda}_{0} \lambda{'} (\lambda_{s} 
	   - \lambda)(\lambda-\lambda{'}) \bigg(\Phi_{00}(\lambda) \Psi_{0}(\lambda{'}) \\
	   &+  \Psi_{0}(\lambda) \Phi_{00}(\lambda{'})\bigg) 
	   d\lambda{'}d\lambda\bigg] 
	   \bar{\zeta}_{o}. 
\end{aligned}
\end{equation}
This is the general solution of the geodesic deviation equation at second order. We proceed now to connect this equation with the lens equation \eqref{eq:0.00}.
In order to do that, we decompose $\Psi_{0}$, $\zeta_{s}$ and $\zeta_{o}$ into their real and imaginary parts,
\begin{equation}
\begin{aligned}
	\Psi_{0}  = \Psi_{0R}  + i \; \Psi_{0I}, \ \
	\zeta_{s} = \zeta_{sR} + i \;\zeta_{sI}, \ \
	\zeta_{o} = \zeta_{oR} + i \;\zeta_{oI}, 
\end{aligned}	
\end{equation}
which after replacing in \eqref{eq:2.41} gives the following equations:
\begin{equation}\label{eq:2777}
\begin{aligned}
	\zeta_{sR} &= A \; \zeta_{oR} + B \; \zeta_{oI}, \\
	\zeta_{sI} &= A{'} \; \zeta_{oR} + B{'} \; \zeta_{oI},
\end{aligned}	
\end{equation}
where the explicit expressions for the coefficients $\{A, B, A{'}, B{'}\}$ can be found in the Appendix~\ref{abcd}.

From the linearity of the lens equation, we know that the deviation vectors must also be related by the same lens mapping matrix $A^i{}_j$~\cite{Uzan:2000xv},
\begin{equation}\label{eq: 2.77}
 \zeta^{i}_{s} = A^{i}{}_{j} \, \zeta^{j}_{o},
\end{equation}
and where $\{\zeta_{s}^{i},\zeta_{o}^{i}\}$ are the spatial vector associated with $\{\zeta_{s},\zeta_{o}\}$. 

Since we are not interested in the component $\zeta_\ell$ of $\zeta^{a}$ along $\ell^{a}$, we will consider only the projection of the deviation vector in the two-space spanned by
$\{ m^{a}, \bar{m}^{a} \}$,
\begin{equation}
	\zeta_{\perp}^{a} = \zeta \bar{m}^{a} + \bar{\zeta} m^{a}.
\end{equation}
Introducing an orthonormal spatial Sachs basis $\{a^{a},b^{a}\}$ at the observer position and by parallel transport of this basis to the other points in the past null cone we can always express $m^{a}$ by
\begin{equation}
	m^{a} = \frac{1}{\sqrt{2}} (a^{a} + i \, b^{a}).
\end{equation}
Hence, we obtain
\begin{equation}
	\zeta_{\perp}^{a} = {\sqrt{2}} (\zeta_{R} a^{a} + \zeta_{I} b^{a}).
\end{equation}
where $\{ \zeta_{R}, \zeta_{I} \}$ are the real and imaginary part of the component $\zeta$, respectively.
Therefore, from \eqref{eq: 2.77} we see that:
\begin{equation}
	\begin{pmatrix}
	\zeta_{sR}     \\   \zeta_{sI} 	
	\end{pmatrix}				
	=
	\begin{pmatrix}
		1 - \kappa - \gamma_1   &   -\gamma_2 - \hat{\omega}   \\
		-\gamma_2 + \hat{\omega}   &   1 - \kappa + \gamma_1   \\
	\end{pmatrix}
	\begin{pmatrix}
	\zeta_{oR}     \\   \zeta_{oI}
	\end{pmatrix}.
\end{equation}
Finally, by comparison with \eqref{eq:2777} we arrive to the following expressions for the optical scalars:

\begin{equation}\label{convergencia:gral2}
\begin{aligned}
\kappa =& \frac {1}{\lambda_{s}} \int^{\lambda_{s}}_{0} \lambda (\lambda_{s} - \lambda) \Phi_{00}(\lambda) d\lambda 
			 - \frac{1}{\lambda_{s}} \int_{0}^{\lambda_{s}}\int_{0}^{\lambda}\lambda{'}(\lambda_{s}-\lambda) \\
			 &(\lambda-\lambda{'}) 
			 \bigg(\Phi_{00}(\lambda)\Phi_{00}(\lambda{'}) 
			 + \Re  \{\Psi_{0}(\lambda) \bar{\Psi}_{0}(\lambda{{'}}) \} \bigg) 
		       	d\lambda{'} d\lambda,
\end{aligned}				
\end{equation}
\begin{equation}\label{shear:gral2}
\begin{aligned}
\gamma &= \frac {1}{\lambda_{s}} \int^{\lambda_{s}}_{0} \lambda (\lambda_{s} - \lambda) \Psi_{0}(\lambda) d\lambda 
				 - \frac{1}{\lambda_{s}} \int_{0}^{\lambda_{s}}\int_{0}^{\lambda}\lambda{'}(\lambda_{s}-\lambda) \\
				 &(\lambda-\lambda{'}) 
				 \bigg( \Phi_{00}(\lambda) \Psi_{0}(\lambda{'}) 
				 + \Psi_{0}(\lambda) \Phi_{00}(\lambda{'}) \bigg) 
				d\lambda{'}d\lambda,
\end{aligned}
\end{equation}
\begin{equation}\label{rotacion:gral2}
\hat{\omega} = \frac{1}{\lambda_{s}} \int_{0}^{\lambda_{s}}\int_{0}^{\lambda}\lambda{'}(\lambda_{s}-\lambda)(\lambda-\lambda{'}) \Im \{ \Psi_{0}(\lambda) \bar{\Psi}_{0}(\lambda{{'}}) \} d\lambda{'}d\lambda;
\end{equation}
where $\Re\{\cdot\}$ and $\Im\{\cdot\}$ indicate real and imaginary part, respectively.

The way in which the different curvature scalars appear in the expression for the convergence $\kappa$ and the shear $\gamma$ is expected taking into account that they are the integrated version of the well-known relations between the local quantities $\rho$ and $\sigma$~\cite{Newman-1962} 
\begin{equation}\label{rho_rho}
\ell(\rho) = (\rho^{2} + \sigma \bar{\sigma}) + \Phi_{00},
\end{equation}
\begin{equation}\label{sigma_sigma}
 \ell(\sigma) = (\rho + \bar{\rho})\sigma + \Psi_{0}.
\end{equation}

On the other hand even when the congruence is twist free,
\begin{equation}
\omega=\frac{1}{2}\left(\nabla_{[a}l_{b]}\nabla^a l^b\right)^{1/2}=\frac{1}{2}(\rho-\bar\rho)=0;
\end{equation}
the rotation of the image described by the scalar $\hat\omega$ could be different from zero due to a cumulative effect of shearing in different directions when the light beams pass different regions of lensing~\cite{Holz-1998}.  

Now, as we are interested in expressions up to second-order in the formalism of weak lenses, we need to address two issues: first, we only need the curvature
components $\Phi_{00}$ and $\Psi_{0}$ up to second-order in the flat metric perturbation; that is
\begin{equation}
\begin{aligned}
 \Phi_{00} &= \Phi_{00}^{(1)} + \Phi_{00}^{(2)} + \mathcal{O}(\varepsilon^{3}), \\
 \Psi_{0} &= \Psi_{0}^{(1)} + \Psi_{0}^{(2)} + \mathcal{O}(\varepsilon^{3}),
\end{aligned}
\end{equation}
where the superscript in parenthesis indicates the order in $\varepsilon$ of the respective quantity, with $\varepsilon$ the parameter which measure the perturbation of a flat metric. Consequently, we need to compute the Ricci and Weyl tensor up to second-order in $\varepsilon$ and perform 
the parallel transport of the null tetrad at first order also in $\varepsilon$. Second, as we are working in the weak lensing regime, we can approximate the actual path of the beam as follows:
\begin{equation}
     x^{a}_{actual}(\lambda) = x^{(0)a}(\lambda) + \delta x^{(1)a}(\lambda) + \mathcal{O}(\varepsilon^{2});
\end{equation}
that is, we can express the actual beam path as its path in the background plus higher order corrections. Here, we only need to consider corrections at first order. This method of approximation which goes beyond of the Born approximation is frequently used in the literature of second order lensing~\cite{1475-7516-2015-06-050,Cooray:2002mj,Bernardeau-2010,Hagstotz:2014qea,Marozzi:2016uob,Petri:2016qya,Schaefer:2005up}.

Therefore, we expand $\Phi_{00}$ and $\Psi_{0}$ along the background geodesic at second-order as, 
\begin{equation}\label{aproxxx}
\begin{aligned}
    \Phi_{00}(x^{a}_{actual}(\lambda)) =& \Phi_{00}^{(1)}(x^{(0)a}(\lambda)) + \delta x^{(1)a}(\lambda) \frac{\partial \Phi_{00}^{(1)}}{\partial x^{a}}\bigg |_{x^{(0)a}(\lambda)} \\
    & + \Phi_{00}^{(2)}(x^{(0)a}(\lambda)), \\
     \Psi_{0}(x^{a}_{actual}(\lambda)) =& \Psi_{0}^{(1)}(x^{(0)a}(\lambda)) + \delta x^{(1)a}(\lambda) \frac{\partial \Psi_{0}^{(1)}}{\partial x^{a}}\bigg |_{x^{(0)a}(\lambda)}\\
    &+ \Psi_{0}^{(2)}(x^{(0)a}(\lambda)).
\end{aligned}
\end{equation}
In conclusion, the final expressions for the optical scalars are formally written in terms of the curvature scalars as
\begin{equation}\label{convergencia:gral}
\begin{aligned}
	\kappa =& \frac {1}{\lambda_{s}} \int^{\lambda_{s}}_{0} \lambda (\lambda_{s} - \lambda) \bigg( \Phi^{(1)}_{00}(\lambda) + \Phi^{(2)}_{00}(\lambda) 
	       + \delta x^{(1)a}(\lambda) \\ 
	       &\frac{\partial \Phi^{(1)}_{00}}{\partial x^{a}}\bigg |_{\lambda}  \bigg) d\lambda 
	       - \frac{1}{\lambda_{s}} \int_{0}^{\lambda_{s}}\int_{0}^{\lambda}\lambda{'}(\lambda_{s}-\lambda)(\lambda-\lambda{'}) \\
	       &\bigg(\Phi^{(1)}_{00}(\lambda)\Phi^{(1)}_{00}(\lambda{'}) 
	       + \Re  \{\Psi_{0}^{(1)}(\lambda) 
	       \bar{\Psi}_{0}^{(1)}(\lambda{{'}}) \} \bigg) d\lambda{'}  d\lambda,
\end{aligned}				
\end{equation}
\begin{equation}\label{shear:gral}
\begin{aligned}
\gamma =& \frac {1}{\lambda_{s}} \int^{\lambda_{s}}_{0} \lambda (\lambda_{s} - \lambda) \bigg( \Psi^{(1)}_{0}(\lambda) + \Psi^{(2)}_{0}(\lambda) 
	+ \delta x^{(1)a}(\lambda) \\
	&\frac{\partial \Psi^{(1)}_{0} }{\partial x^{a}}\bigg |_{\lambda} \bigg) d\lambda 
	- \frac{1}{\lambda_{s}} \int_{0}^{\lambda_{s}}\int_{0}^{\lambda}\lambda{'}(\lambda_{s}-\lambda)(\lambda-\lambda{'}) \\
	&\bigg( \Phi^{(1)}_{00}(\lambda) \Psi^{(1)}_{0}(\lambda{'}) + \Psi^{(1)}_{0}(\lambda) \Phi^{(1)}_{00}(\lambda{'}) \bigg) 
	d\lambda{'}d\lambda,
\end{aligned}
\end{equation}
\begin{equation}\label{rotacion:gral}
\hat{\omega} = \frac{1}{\lambda_{s}} \int_{0}^{\lambda_{s}}\int_{0}^{\lambda}\lambda{'}(\lambda_{s}-\lambda)(\lambda-\lambda{'})  \Im  \{ \Psi_{0}^{(1)}(\lambda) \bar{\Psi}_{0}^{(1)}(\lambda{'}) \} d\lambda{'}d\lambda.
\end{equation}

Equations (\ref{convergencia:gral}), (\ref{shear:gral}) and (\ref{rotacion:gral}) are the most general formal expressions for the optical scalars which can be obtained
without mentioning neither the extension of the lens nor their shape.
We want to emphasize that unlike the expressions (\ref{convergencia:gral2}), (\ref{shear:gral2}), (\ref{rotacion:gral2}),
in these last expressions the integrals are made over the background geodesic and only contain quantities up to second-order.
These formulas for the optical scalars generalize to second order the relations found in~\cite{PhysRevD.83.083007} (see also\cite{0264-9381-22-14-005}).

\section{The thin lens approximation}\label{section4}
\subsection{Preliminary considerations}
As is well-known, the thin lens approximation is based  on the  assumption that the \emph{lens-observer} and \emph{lens-source} distances are significantly larger that the lens size
~\citep{Ehlers:1992dau}. In this situation, the different optical scalars can be found by projection of the curvature scalars in the line of sight~\cite{0264-9381-22-14-005,Seitz:1994xf}.

Let $C$ or $D$ be any of the $\{ \Phi_{00}, \Psi_{0} \}$ scalars so, following the approach used in \cite{PhysRevD.83.083007} the first order thin lens approximation is completely contained
in the following expression
\begin{equation}\label{aprox-delgada}
	\widetilde{C}(\lambda{}) \equiv \int^{\lambda{}}_{0} C(\lambda{'}) d\lambda{'} \cong \left\{
	       \begin{array}{ll}
		 0       & \forall \ \lambda{'} < \lambda_{l} - \delta \\
		 \hat{C} & \forall \ \lambda{'} \geq \lambda_{l} + \delta \\
	       \end{array}
	     \right.
\end{equation}
where $\delta \ll \lambda_{l}, \delta \ll \lambda_{ls}, \delta \ll \lambda_{s}$ ($\lambda_{ls} \equiv \lambda_{s}-\lambda_{l}$). 

On the other hand, even when $\delta \ll \lambda_{l}$, $\delta \ll \lambda_{ls}$ implies that \eqref{aprox-delgada} is a good approximation in the computation of first order quantities, it does not mean that it remains sufficiently precise in order to compute the second order quadratic terms. We prove in the Appendix \ref{app-thin-lens}, that if we only consider the approximation \eqref{aprox-delgada} in the computation of quantities like 
\begin{equation}\label{terminos_cuadraticos_tipo}
   \int_{0}^{\lambda_{s}}\int_{0}^{\lambda}\lambda{'}(\lambda_{s}-\lambda)(\lambda-\lambda{'})
    C(\lambda{}) D(\lambda{'}) d\lambda{'} d\lambda{};
\end{equation}
then these quantities should be zero. However, even for a point mass lens, as we will show in Sec.\ref{Applications} terms like \eqref{terminos_cuadraticos_tipo} make a non-negligible contribution. It should not be seen as a surprising result, for example even in a toy model where quantities like $C(\lambda)$ are represented by step functions centered in $\lambda_l$ and with a width $\delta$, integrals like \eqref{terminos_cuadraticos_tipo} make nontrivial contributions proportional to the width $\delta$. Unfortunately, even when in the thin lens situation we can find more simple expressions for the integrals which depend linearly from the curvature scalars (they can be written in terms of projected $\hat{C}$ quantities), we can not do the same with the quadratic ones. Therefore, in the following we only express the linear integrals in $C$ in terms if $\hat{C}$ and rewrite in a more compact way the integrals which are quadratic in the curvature scalars. 

In order to implement the thin lens approximation to the optical scalars \eqref{convergencia:gral2}, \eqref{shear:gral2}, \eqref{rotacion:gral2},
we need to manipulate these expressions as follows.
Using (\ref{aprox-delgada}) in (\ref{convergencia:gral2}) and (\ref{shear:gral2}) we have terms of the form
\begin{equation}\label{a100}
\begin{aligned}
	\int_{0}^{\lambda_s} \lambda (\lambda_{s}-\lambda) C(\lambda) d\lambda 
	= \hat{C} \lambda_{l} \lambda_{ls}.
\end{aligned}
\end{equation}
That is, the integrals which are linear in the curvature scalars $C$ can be written in a more compact way in terms of the projected quantities $\hat{C}$~\cite{PhysRevD.83.083007,0264-9381-22-14-005}.

On the other hand, even when we can not do the same with the terms which are quadratic in $C$ and $D$, we use the following notation which will be useful later. The contribution of these terms to the optical scalars will be denoted as
\begin{equation}
\begin{aligned}
\left[\cdot\right]^{(2)}_{CD}=&-\frac{1}{\lambda_s}\int_{0}^{\lambda_{s}}\int_{0}^{\lambda}\lambda{'}(\lambda_{s}-\lambda)(\lambda-\lambda{'}) \\ 
    & \times C(\lambda{}) D(\lambda{'}) d\lambda{'}d\lambda{};
    \end{aligned}
\end{equation}
where into the squares brackets $\left[\cdot\right]$ should be placed the associated quantity to the respective optical scalar. Therefore, we have quadratic contributions to the second order optical scalars with terms like 
${\kappa}^{(2)}_{\text{\tiny $\Phi$}\text{\tiny $\Phi$}}$, ${\kappa}^{(2)}_{\text{\tiny $\Psi$}\text{\tiny $\bar\Psi$}}$, ${\gamma}^{(2)}_{\text{\tiny $\Phi$}\text{\tiny $\Psi$}}$ and ${\gamma}^{(2)}_{\text{\tiny $\Psi$}\text{\tiny $\Phi$}}$.

Multiplying each of them by the prefactor $\frac{\lambda_s}{\lambda_l\lambda_{ls}}$, we define the associated scalars $\tilde{\kappa}^{(2)}_{\text{\tiny $\Phi$}\text{\tiny $\Phi$}}$, $\tilde{\kappa}^{(2)}_{\text{\tiny $\Psi$}\text{\tiny $\bar\Psi$}}$, $\tilde{\gamma}^{(2)}_{\text{\tiny $\Phi$}\text{\tiny $\Psi$}}$ and $\tilde{\gamma}^{(2)}_{\text{\tiny $\Psi$}\text{\tiny $\Phi$}}$ by:
\begin{equation}\label{tildeka}
\begin{aligned}
\tilde{\kappa}^{(2)}_{\text{\tiny $\Phi$}\text{\tiny $\Phi$}}=\frac{\lambda_s}{\lambda_l\lambda_{ls}}{\kappa}^{(2)}_{\text{\tiny $\Phi$}\text{\tiny $\Phi$}}=
-\frac{1}{\lambda_l\lambda_{ls}}\int_{0}^{\lambda_{s}}\int_{0}^{\lambda}\lambda{'}(\lambda_{s}-\lambda)(\lambda-\lambda{'}) \\
\times \;  \Phi_{00}(\lambda{'}) \Phi_{00}(\lambda{})d\lambda{'} d\lambda;&
 \end{aligned}
\end{equation}
with similar relations for the rest of the quadratic terms. 
Finally, using the approximation \eqref{aproxxx}  we obtain for the convergence and shear
\begin{equation}\label{eq:2.34}
\begin{aligned}
    \kappa &= \frac{\lambda_{l} \lambda_{ls}}{\lambda_{s}} \bigg( \hat{\Phi}_{00}^{(1)} + \delta \hat{\Phi}_{00}^{(2)} +\hat{\Phi}_{00}^{(2)}+\tilde{\kappa}^{(2)}_{\text{\tiny $\Phi$}\text{\tiny $\Phi$}}+\Re [\tilde{\kappa}^{(2)}_{\text{\tiny $\Psi$}\text{\tiny $\bar\Psi$}}] \bigg),  
\end{aligned}
\end{equation}
 \begin{equation} \label{eq:2.33}
\begin{aligned}
	&\gamma = \frac{\lambda_{l} \lambda_{ls}}{\lambda_{s}} \bigg( \hat{\Psi}_{0}^{(1)} + \delta \hat{\Psi}_{0}^{(2)} +\hat{\Psi}_{0}^{(2)} + \tilde{\gamma}^{(2)}_{\text{\tiny $\Phi$}\text{\tiny $\Psi$}} +\tilde{\gamma}^{(2)}_{\text{\tiny $\Psi$}\text{\tiny $\Phi$}}\bigg).
\end{aligned}
\end{equation}

As $\hat\omega$ does not have linear terms in the curvature, it remains the same as in \eqref{rotacion:gral}.
Note that in the previous expressions, the projected  hat quantities are given by 
\begin{equation}
\begin{aligned}
	\hat{\Phi}_{00}^{(j)} =& \int^{\lambda_{s}}_{0} \Phi_{00}^{(j)} d\lambda, \ \ \text{$j = 1, 2$},\\
    \hat{\Psi}_{0}^{(j)}=& \int^{\lambda_{s}}_{0} \Psi_{0}^{(j)} d\lambda, \ \ \text{$j = 1, 2$},\\
	\delta \hat{\Phi}_{00}^{(2)} =& \int^{\lambda_{s}}_{0} \delta x^{(1)a}(\lambda) \frac{\partial \Phi_{00}^{(1)}}{\partial x^{a}}\bigg|_{x^{(0)}(\lambda)} d\lambda, \\
	\delta \hat{\Psi}_{0}^{(2)} =& \int^{\lambda_{s}}_{0} \delta x^{(1)a}(\lambda) \frac{\partial \Psi_{0}^{(1)}}{\partial x^{a}}\bigg|_{x^{(0)}(\lambda)} d\lambda.
\end{aligned}
\end{equation}

\subsection{Axisymmetric lenses}
Let us consider a static and axisymmetric gravitational lens in the thin lens approximation where the axis of symmetry corresponds to
the line of sight which crosses the central region of the matter distribution from the observer. 
We select a Cartesian coordinate system with the origin in the lens plane
and the line of sight in the negative $y$ direction. As in~\cite{PhysRevD.83.083007}, in the lens plane we identify the first component with the $z$ coordinate and the second component with 
the $x$ coordinate. In this plane, it is convenient to work with new coordinates $(J,\vartheta)$ given by
\begin{equation}\label{eq:a.01}
\begin{aligned}
      z &= J \cos(\vartheta), \\
      x &= J \sin(\vartheta), 
\end{aligned}
\end{equation}
where $J$ is the impact parameter and $\vartheta$ is the polar angle measured from $z$.

Since $\Phi_{00}$ and $\Psi_{0}$ are quantities with spin weight 0 and 2 respectively, we assume that in the case of static and axial symmetries they have 
the following functional dependence\cite{0264-9381-22-14-005}:
\begin{equation}\label{psipsi}
 \begin{aligned}
    \Phi_{00} &= \Phi_{00}(y,J), \\
    \Psi_{0} &= -\psi_{0}(y,J) e^{2i\vartheta}, 
 \end{aligned}
\end{equation}
for some arbitrary phase in the choice of $m^{a}$. Then, the projected curvature scalars are given by
\begin{equation}\label{hatspherically}
 \begin{aligned}
    \hat{\Phi}_{00} &= \hat{\Phi}_{00}(J), \\
    \hat{\Psi}_{0} &= -\hat{\psi}_{0}(J) e^{2i\vartheta} ,
 \end{aligned}
\end{equation}
where
\begin{equation}
    \hat{\psi}_{0}(J) = -e^{-2i\vartheta} \int^{\lambda_{s}}_{0} \Psi_{0}(\lambda) d\lambda.
\end{equation}
According to the expressions \eqref{psipsi} and \eqref{hatspherically} we can write the optical scalars \eqref{eq:2.34}, \eqref{eq:2.33} as follows:
\begin{equation}\label{kappathin}
\begin{aligned}
    \kappa =& \frac{\lambda_{l} \lambda_{ls}}{\lambda_{s}} \bigg( \hat{\Phi}_{00}^{(1)}(J) + \delta \hat{\Phi}_{00}^{(2)}(J) + \hat{\Phi}_{00}^{(2)}(J)\\ &+\tilde{\kappa}^{(2)}_{\text{\tiny $\Phi$}\text{\tiny $\Phi$}}(J,\lambda_l,\lambda_{ls})+\Re [\tilde{\kappa}^{(2)}_{\text{\tiny $\Psi$}\text{\tiny $\bar\Psi$}}(J,\lambda_l,\lambda_{ls})]\bigg) ,
 \end{aligned}
\end{equation}
\begin{equation}\label{gamma1thin}
\begin{aligned}
	\gamma_{1} =& - \frac{\lambda_{l} \lambda_{ls}}{\lambda_{s}} \bigg( \hat{\psi}_{0}^{(1)}(J) + \delta \hat{\psi}_{0}^{(2)}(J) + \hat{\psi}_{0}^{(2)}(J)\bigg.\\
  &+\bigg.\underline{\tilde{\gamma}^{(2)}_{\text{\tiny $\Phi$}\text{\tiny $\Psi$}}}(J,\lambda_l,\lambda_{ls}) +\underline{\tilde{\gamma}^{(2)}_{\text{\tiny $\Psi$}\text{\tiny $\Phi$}}}(J,\lambda_l,\lambda_{ls})\bigg)
	 \cos(2\vartheta),
\end{aligned}
\end{equation}
\begin{equation}\label{gamma2thin}
\begin{aligned}
	\gamma_{2} =& - \frac{\lambda_{l} \lambda_{ls}}{\lambda_{s}} \bigg( \hat{\psi}_{0}^{(1)}(J) + \delta \hat{\psi}_{0}^{(2)}(J) + \hat{\psi}_{0}^{(2)}(J)\bigg.\\
  &+\bigg.\underline{\tilde{\gamma}^{(2)}_{\text{\tiny $\Phi$}\text{\tiny $\Psi$}}}(J,\lambda_l,\lambda_{ls}) +\underline{\tilde{\gamma}^{(2)}_{\text{\tiny $\Psi$}\text{\tiny $\Phi$}}}(J,\lambda_l,\lambda_{ls}) \bigg) 
	 \sin(2\vartheta),
\end{aligned}
\end{equation}
where 
\begin{equation}
\hat{\psi}_{0}^{(j)}= \int^{\lambda_{s}}_{0} \psi_{0}^{(j)} d\lambda, \ \ \text{$j = 1, 2$},
\end{equation}
\begin{equation}
 \delta \hat{\psi}_{0}^{(2)} = \int^{\lambda_{s}}_{0} \delta x^{(1)a}(\lambda) \frac{\partial \psi_{0}^{(1)}}{\partial x^{a}}\bigg|_{x^{(0)}(\lambda)} d\lambda,
\end{equation}
and
\begin{equation}
\begin{aligned}
\underline{\tilde{\gamma}^{(2)}_{\text{\tiny $\Phi$}\text{\tiny $\Psi$}}}=& -\frac{1}{\lambda_{l}\lambda_{ls}} \int_{0}^{\lambda_{s}}\int_{0}^{\lambda}\lambda{'}(\lambda_{s}-\lambda)(\lambda-\lambda{'}) 
\psi_{0}(\lambda,J) \\ &\Phi_{00}(\lambda{'},J) d\lambda{'} d\lambda;
\end{aligned} 
\end{equation}
with a similar definition for $\underline{\tilde{\gamma}^{(2)}_{\text{\tiny $\Psi$}\text{\tiny $\Phi$}}}$.

\subsection{Deflection angle}

Now, we want to obtain an expression for the deflection angle in the thin lens approximation which generalizes \eqref{angulito}. 

From the lens equation \eqref{eq: 0.1}
 we see that in this approximation the amplification matrix $A^i{}_j$ can be expressed as~\cite{Bartelmann:2010fz,Narayan:1996ba}
\begin{equation}\label{eq:thin}
      A^{i}{}_{j} = \frac{d\beta^{i}}{d\theta^{j}} 
      = \delta^{i}{}_{j} - \frac{\lambda_{ls}\lambda_{l}}{\lambda_{s}} \frac{d\alpha^{i}}{dx^{j}},
\end{equation}
where we have used that in the thin lens approximation $\frac{d}{d\theta^{i}} \approx \lambda_{l}\frac{d}{d x^{i}}$. We define the components of 
 $\alpha^{i} = (\alpha^{1},\alpha^{2})$ as
\begin{equation}\label{eq:alphai}
      \alpha^{i} = \alpha(J,\lambda_l,\lambda_{ls})\bigg(\frac{z}{J},\frac{x}{J} \bigg).
\end{equation}

Hence, from (\ref{eq:thin}), \eqref{eq:alphai} and  \eqref{Aij} follow that 
\begin{equation}\label{eq:2.66}
      \kappa - \gamma_{1} \cos(2\vartheta) - \gamma_{2} \sin(2\vartheta) = \frac{\lambda_{ls}\lambda_{l}}{\lambda_{s}} \frac{\alpha}{J}.
\end{equation}

Therefore, since we are interested in the deflection angle in the asymptotic region, that is where $\lambda_l$ and $\lambda_{ls}$ go to infinity, from (\ref{eq:2.66}) and the expressions \eqref{kappathin}, \eqref{gamma1thin}, \eqref{gamma2thin} for the optical scalars, we finally obtain
\begin{widetext}
\begin{equation}\label{eqf02}
\begin{aligned}
       \alpha(J) &= J \bigg( \hat{\Phi}^{(1)}_{00}(J) + \hat{\psi}^{(1)}_{0}(J) + \delta \hat{\Phi}^{(2)}_{00}(J) +  \delta \hat{\psi}^{(2)}_{0}(J) 
       + \hat{\Phi}^{(2)}_{00}(J) + \hat{\psi}^{(2)}_{0}(J)+\tilde{\kappa}^{(2)}_{\text{\tiny $\Phi$}\text{\tiny $\Phi$}}(J)+\Re [\tilde{\kappa}^{(2)}_{\text{\tiny $\Psi$}\text{\tiny $\bar\Psi$}}(J)]+\underline{\tilde{\gamma}^{(2)}_{\text{\tiny $\Phi$}\text{\tiny $\Psi$}}}(J) +\underline{\tilde{\gamma}^{(2)}_{\text{\tiny $\Psi$}\text{\tiny $\Phi$}}}(J)  \bigg);
\end{aligned}
\end{equation}
\end{widetext}
where 
\begin{equation}
\tilde{\kappa}^{(2)}_{\text{\tiny $\Phi$}\text{\tiny $\Phi$}}(J):=\lim_{\lambda_l,\lambda_{ls}\to\infty}\tilde{\kappa}^{(2)}_{\text{\tiny $\Phi$}\text{\tiny $\Phi$}}(J,\lambda_l,\lambda_{ls}),
\end{equation}
and with similar definitions for the other  quadratic terms.

This expression generalizes our previous formula \eqref{angulito} to second order. At the difference of similar relations which are written in terms of metric components, this quantity is explicitly gauge invariant, due to it only depends on well-defined quantities as the impact parameter $J$ and curvature scalars. We will see in the examples of the next section how using this formula to compute the deflection angle at second order in two different coordinate systems yields the same result.
\section{Applications}\label{Applications}
\subsection{Optical scalars for a Schwarzschild metric at second order}

As a first example we will compute explicitly the optical scalars and deflection angle for a  Schwarzschild point mass lens in two different coordinate systems. 

For this, we proceed as follows. First, we choose a null tetrad such that far away from the point mass, in the asymptotically flat region it reduces to
\begin{equation}
\begin{aligned}
 \ell^{a} &= (-1,0,1,0), \ \ \ \ \  m^{a} = \frac{1}{\sqrt{2}} (0,i,0,1), \\
 n^{a} &= \frac{1}{2} (-1,0,-1,0), \ \bar{m}^{a} = \frac{1}{\sqrt{2}} (0,-i,0,1).
\end{aligned}
\end{equation}

On the other hand, we choose the origin of the coordinate system in the lens's position and parametrize the geodesic by
\begin{equation}
 ( x(\lambda), y(\lambda), z(\lambda) ) = ( x, \lambda - \lambda_{l}, z),
\end{equation}
that is, $\lambda = 0$ indicates the observer's position and $\lambda = \lambda_{s}$ the source's position at $\lambda_{ls}$. Without loss of generality we can take $x=J$ and $z=0$ (and therefore $\vartheta=\frac{\pi}{2}$).
In order to analyze how the different aspects of the curvature contribute to the optical scalars and taking into account the Eq.\eqref{tildeka} we define
\begin{eqnarray}\label{eq:redef}
\tilde\kappa&:=&\frac{\lambda_s}{\lambda_{ls}\lambda_l}\kappa,\\
\tilde\gamma&:=&\frac{\lambda_s}{\lambda_{ls}\lambda_l}\gamma.
\end{eqnarray}

\subsubsection{Isotropic coordinates}
Let us consider a static, spherically symmetric body acting as a gravitational lens in such a way that its external gravitational field can be described by the Schwarzschild metric.
This metric can be expressed in isotropic coordinates by
\begin{equation}\label{isotropic}
 ds^{2} = \frac{\bigg(1 - \frac{\varepsilon}{4 r}\bigg)^{2}}{\bigg(1 + \frac{\varepsilon}{4 r}\bigg)^{2}} dt^{2} - 
        \bigg(1 + \frac{\varepsilon}{4 r}\bigg)^{4} (dx^{2} + dy^{2} + dz^{2}),
\end{equation}
where $\varepsilon = 2 M $ and $r = \sqrt{x^{2}+y^{2}+z^{2}}$.

The corresponding null tetrad, curvature scalars and the first order correction to the geodesic are shown in the Appendix~\ref{A:sch}. The integrals needed in the computation of the optical scalars were made with the help of MAPLE and the GRTensor package. 

The resulting leading order behavior for the  optical scalars is given by
\begin{equation}\label{schw-gamma}
\begin{aligned}
 \tilde{\gamma}(J) =& \bigg[\frac{2}{J^2} + \mathcal{O}(\frac{1}{\lambda_{ls}},\frac{1}{\lambda_{l}}) \bigg] \varepsilon 
	    + \bigg[ \frac{45 \pi}{32 J^{3}} + \mathcal{O}(\frac{1}{\lambda_{ls}},\frac{1}{\lambda_{l}}) \bigg] \varepsilon^{2}\\& + \mathcal{O}(\varepsilon^{3});
\end{aligned}	    
\end{equation}
\begin{equation}\label{schw-kappa}
\begin{aligned}
  \tilde{\kappa}(J) &= \bigg[ - \frac{15 \pi}{32 J^{3}}  + \mathcal{O}(\frac{1}{\lambda_{ls}},\frac{1}{\lambda_{l}}) \bigg] \varepsilon^{2} 
		 + \mathcal{O}(\varepsilon^{3}),
\end{aligned}
\end{equation}
while $\hat{\omega} = 0$.

If we take the limits $\lambda_{l} \to \infty $ and $ \lambda_{ls} \to \infty$ of these relations and using \eqref{eqf02} [or equivalently \eqref{eq:2.66}], we obtain an expression for the deflection angle which agrees with
the familiar result for the bending angle at second-order for the Schwarzschild metric
\begin{equation}\label{eq:Salpha}
 \alpha(J) = \frac{4\,M}{J} + \frac{15\,\pi}{4} \frac{M^{2}}{J^{2}}.
\end{equation}
\subsubsection{Quasi-Minkowskian coordinates}
It is important to note that, the expressions \eqref{schw-gamma}, \eqref{schw-kappa} and \eqref{eq:Salpha} are coordinate independent because 
they only depend on the impact parameter $J$ and the total mass $M$. 
We could make similar computations for the same metric in a different gauge. For example, using the so-called quasi-Minkowskian coordinate system where the Schwarzschild metric reads\cite{Weinberg-1973}: 
\begin{equation}\label{cartesian-coord}
\begin{aligned}
 ds^{2} = &\bigg(1 - \frac{\varepsilon}{r} \bigg) dt^{2} - (dx^{2} + dy^{2} + dz^{2}) \\
 &- \bigg[ \bigg( 1 - \frac{\varepsilon}{r} \bigg)^{-1} - 1 \bigg] r^{-2} ( x dx + y dy + z dz )^{2},
\end{aligned} 
\end{equation}
where again $\varepsilon = 2M$ and $r=\sqrt{x^{2}+y^{2}+z^{2}}$, we obtain the same results for the optical quantities.
(For a complete discussion and more details, see table \ref{T1} and Appendix \ref{A:sch}).

\subsection{Parametrized-Post-Newtonian point mass lens}
We will consider now a more general metric known as the \emph{Parametrized-post-Newtonian (PPN) point mass metric} whose line element is given by
\begin{equation}
 ds^{2} = \bigg( 1 - \frac{\varepsilon}{r} + \frac{\beta \varepsilon^{2}}{2 r^{2}} \bigg) dt^{2} - \bigg( 1 + \frac{\mu \varepsilon}{r} + \frac{3 \nu \varepsilon^{2}}{8 r^{2}} \bigg)
	(dx^{2} + dy^{2} + dz^{2}),
\end{equation}
where $\varepsilon = 2 M $ and $r = \sqrt{x^{2}+y^{2}+z^{2}}$.
For the particular choice  $\beta = \mu = \nu = 1$, this metric reduces to the second order approximation of the Schwarzschild metric in isotropic coordinates. In the Appendix \ref{A:sch} are described the corresponding null tetrad, first order correction to the geodesic and curvature scalars. 

The leading order behavior for the optical scalars in the PPN metric is given by 
\begin{equation}
 \tilde{\kappa} = \bigg[\frac{\pi(-8+4\beta-3\nu-8\mu)}{32J^{3}} + \mathcal{O}(\frac{1}{\lambda_{ls}},\frac{1}{\lambda_{l}})\bigg]\varepsilon^{2}+ \mathcal{O}(\varepsilon^{3}),
\end{equation}
\begin{equation}
\begin{aligned}
 \tilde{\gamma} =& \bigg[\frac{(1+\mu)}{J^2}+\mathcal{O}(\frac{1}{\lambda_{ls}},\frac{1}{\lambda_{l}})\bigg]\varepsilon - \bigg[\frac{3\pi(-8+4\beta-3\nu-8\mu)}{32J^{3}} \\ &+\mathcal{O}(\frac{1}{\lambda_{ls}},\frac{1}{\lambda_{l}})\bigg]\varepsilon^{2}+ \mathcal{O}(\varepsilon^{3}).
\end{aligned} 
\end{equation}
Again, if we take the limits $\lambda_{l} \to \infty $ and $ \lambda_{ls} \to \infty$ and using \eqref{eqf02} we recover the well-known result for the deflection angle\cite{Epstein-1980} 
\begin{equation}
\alpha(J)=2(1+\mu)\frac{M}{J} + \pi(2 - \beta  + 2 \mu + \frac{3}{4}\nu)\frac{M^{2}}{J^{2}}.
\end{equation}
\begin{table*}
	\caption{Leading order contribution to the convergence and shear due to the different components of the curvature. Each of the tilde quantities is obtained from the corresponding dagger quantities multiplying by a factor $(\lambda_l\lambda_{ls})^{-1}$. In the case of the Schwarzschild solution, note that even when the contribution to $\tilde{\gamma}$ from the terms $\tilde{\gamma}_{\text{\tiny $\Psi^{(2)}$}}$ and $\tilde{\gamma}_{\text{\tiny $\delta$}\text{\tiny $\Psi$}}$ is gauge dependent, the total contribution in the asymptotic region, is the same.} 
	\label{T1}
	\begin{ruledtabular}
		\begin{tabular}{lccc}
			Optical scalars &Isotropic\footnote{Schwarzschild in isotropic coordinates}		 & q-Minkowskian\footnote{Schwarzschild in quasi-Minkoskian coordinates}	  		  & PPN\footnote{Parametrized-post-Newtonian point mass metric} \\
			\hline 		 
			$\tilde{\kappa}_{\text{\tiny $\Phi^{(1)}$}}$ & 0 		                         & 0 			 		  & 0									\\ 
			$\tilde{\kappa}_{\text{\tiny $\Phi^{(2)}$}}$ & 0                                      & 0					  & $\frac{\pi(-13+8\beta-6\nu+2\mu+9\mu^{2})}{16} \frac{M^{2}}{J^{3}}$	\\ 
			$\tilde{\kappa}_{\text{\tiny $\delta$}\text{\tiny $\Phi$}}$ & 0				         & 0					  & $-\frac{\pi(-2+\mu+\mu^{2})}{8} \frac{M^{2}}{J^{3}}$		\\ 
			$\tilde{\kappa}_{\text{\tiny $\Phi$}\text{\tiny $\Phi$}}$ & 0 					 & 0	   				  & $\frac{\pi(1-2\mu+\mu^{2})}{32} \frac{M^{2}}{J^{3}}$		\\ 
			$\tilde{\kappa}_{\text{\tiny $\Psi$}\text{\tiny $\Psi$}}$ & $-\frac{15\pi}{8} \frac{M^{2}}{J^{3}}$ & $-\frac{15\pi}{8} \frac{M^{2}}{J^{3}}$ & $-\frac{15\pi(1+2\mu+\mu^{2})}{32} \frac{M^{2}}{J^{3}}$		\\ 
			
			\hline\\
			
			$\tilde{\kappa}$ (Total contribution at second order) & $-\frac{15\pi}{8} \frac{M^{2}}{J^{3}}$ & $-\frac{15\pi}{8} \frac{M^{2}}{J^{3}}$ & $\frac{\pi(-8+4\beta-3\nu-8\mu)}{8}\frac{M^2}{J^{3}}$		\\
			\\
			\hline\hline \\
			$\tilde{\gamma}_{\text{\tiny $\Psi^{(1)}$}}$ & $\frac{4M}{J^2}$ 		                   & $\frac{4M}{J^2}$		                 & $2(1+\mu) \frac{M}{J^2}$	\\ 
			
			$\tilde{\gamma}_{\text{\tiny $\Psi^{(2)}$}}$ & $\frac{3\pi}{8} \frac{M^{2}}{J^{3}}$       & $\frac{15\pi}{4} \frac{M^{2}}{J^{3}}$	 &  $\frac{\pi(45-24\beta+18\nu+6\mu-39\mu^{2})}{16} \frac{M^{2}}{J^{3}}$	\\ 
			$\tilde{\gamma}_{\text{\tiny $\delta$}\text{\tiny $\Psi$}}$ & $\frac{21\pi}{4} \frac{M^{2}}{J^{3}}$	   & $\frac{15\pi}{8} \frac{M^{2}}{J^{3}}$	 & $\frac{3\pi(2+7\mu+5\mu^{2})}{8} \frac{M^{2}}{J^{3}}$	\\ 
			$\tilde{\gamma}_{\text{\tiny $\Phi$}\text{\tiny $\Psi$}}$ & 0 					           & 0	   				 	         & $\frac{9\pi(\mu^{2}-1)}{32} \frac{M^{2}}{J^{3}}$	\\ 
			$\tilde{\gamma}_{\text{\tiny $\Psi$}\text{\tiny $\Phi$}}$ & 0     					   & 0 							 & $\frac{9\pi(\mu^{2}-1)}{32} \frac{M^{2}}{J^{3}}$ \\
			
			\hline\\
			
			$\tilde{\gamma}$ (Total contribution at second order) &  $\frac{4M}{J^2}  
			+ \frac{45\pi}{8}\frac{M^{2}}{J^{3}}$    					   & 			$\frac{4M}{J^2}  
			+ \frac{45\pi}{8}\frac{M^{2}}{J^{3}}$			 &  $2{(1+\mu)}\frac{M}{J^2} + \frac{3\pi(8-4\beta+3\nu+8\mu)}{8}\frac{M^2}{J^{3}}$ \\
			\\
		\end{tabular}
	\end{ruledtabular}
\end{table*}

\begin{widetext}
\subsection{Term by term contribution}
Using the definitions~\eqref{eq:redef}, we denote the different
terms in the expressions of the optical scalars as follows:

\begin{equation}
\begin{aligned}
  \tilde{\kappa} =& \frac{1}{\lambda_{ls} \lambda_{l}} \bigg[ \underbrace{ \int_{0}^{\lambda_{s}} \lambda (\lambda_{s}-\lambda) \Phi_{00}^{(1)}(\lambda) d\lambda}_{\text{\large ${\kappa}^\dagger_{\text{\tiny $\Phi^{(1)}$}}$}}
		 + \underbrace{ \int_{0}^{\lambda_{s}} \lambda (\lambda_{s}-\lambda) \Phi_{00}^{(2)}(\lambda) d\lambda}_{\text{\large ${\kappa}^\dagger_{\text{\tiny $\Phi^{(2)}$}}$}} 
		 + \underbrace{ \int_{0}^{\lambda_{s}} \lambda (\lambda_{s}-\lambda) \delta x^{(1)a}(\lambda)
		 \frac{\partial\Phi_{00}^{(1)}}{\partial x^{a}}\bigg |_{\lambda} d\lambda}_{\text{\large ${\kappa}^\dagger_{\text{\tiny $\delta\Phi$}}$}} \\
		 &\underbrace{-  \int_{0}^{\lambda_{s}} \int_{0}^{\lambda^{}} \lambda^{'}(\lambda_{s}-\lambda)(\lambda-\lambda^{'})
		 \Phi_{00}^{(1)}(\lambda) \Phi_{00}^{(1)}(\lambda^{'}) d\lambda^{'} d\lambda}_{\text{\large ${\kappa}^\dagger_{\text{\tiny $\Phi\Phi$}}$}}
		\underbrace{-  \int_{0}^{\lambda_{s}} \int_{0}^{\lambda^{}} \lambda^{'}(\lambda_{s}-\lambda)(\lambda-\lambda^{'})
		 \Re \{\Psi_{0}^{(1)}(\lambda) \Psi_{0}^{(1)}(\lambda^{'})\} d\lambda^{'} d\lambda}_{\text{\large ${\kappa}^\dagger_{\text{\tiny $\Psi\Psi$}}$}} \bigg];
\end{aligned}
\end{equation}
\begin{equation}
\begin{aligned}
  \tilde{\gamma} =& \frac{1}{\lambda_{ls} \lambda_{l}} \bigg[ \underbrace{ \int_{0}^{\lambda_{s}} \lambda (\lambda_{s}-\lambda) \Psi_{0}^{(1)}(\lambda) d\lambda}_{\text{\large ${\gamma}^\dagger_{\text{\tiny $\Psi^{(1)}$}}$}}
		 + \underbrace{ \int_{0}^{\lambda_{s}} \lambda (\lambda_{s}-\lambda) \Psi_{0}^{(2)}(\lambda) d\lambda}_{\text{\large ${\gamma}^\dagger_{\text{\tiny $\Psi^{(2)}$}}$}} 
		 + \underbrace{ \int_{0}^{\lambda_{s}} \lambda (\lambda_{s}-\lambda) \delta x^{(1)a}(\lambda)
		 \frac{\partial\Psi_{0}^{(1)}}{\partial x^{a}}\bigg |_{\lambda} d\lambda}_{\text{\large ${\gamma}^\dagger_{\text{\tiny $\delta\Psi$}}$}} \\
		 &\underbrace{-  \int_{0}^{\lambda_{s}} \int_{0}^{\lambda^{}} \lambda^{'}(\lambda_{s}-\lambda)(\lambda-\lambda^{'})
		 \Phi_{00}^{(1)}(\lambda) \Psi_{0}^{(1)}(\lambda^{'}) d\lambda^{'} d\lambda}_{\text{\large ${\gamma}^\dagger_{\text{\tiny $\Phi\Psi$}}$}} 
		 \underbrace{-  \int_{0}^{\lambda_{s}} \int_{0}^{\lambda^{}} \lambda^{'}(\lambda_{s}-\lambda)(\lambda-\lambda^{'})
		 \Psi_{0}^{(1)}(\lambda) \Phi_{00}^{(1)}(\lambda^{'}) d\lambda^{'} d\lambda}_{\text{\large ${\gamma}^\dagger_{\text{\tiny $\Psi\Phi$}}$}} \bigg].
\end{aligned}
\end{equation}	
\end{widetext}

Using these definitions, in Table \ref{T1} we show how each of these terms contribute to the optical scalars. In the case of the Schwarzschild metric we obtain unequal expressions for $\tilde{\gamma}_{\text{\tiny $\Psi^{(2)}$}}$ and $\tilde{\gamma}_{\text{\tiny $\delta$}\text{\tiny $\Psi$}}$ when the metric is expressed in two different coordinate systems. It is not unexpected because we are considering the same geodesic but in two different coordinate systems. However, their addition contribute in the same way to the total shear. Note also that at second order, the convergence is different from zero as a consequence of the Weyl-Weyl interaction. It means, that even when the Schwarzschild solution can be though as a thin lens, there exist nontrivial contributions to the convergence which comes from interactions between the curvature components in closely but different regions.

In the case of PPN metrics the contribution to the convergence from the term $\tilde{\kappa}_{\text{\tiny $\Phi^{(2)}$}}$ is different from zero, due to the fact that the Ricci scalar $\Phi_{00}$ for this general family of metrics is not zero. Despite that, in the thin lens approximation $\tilde{\kappa}_{\text{\tiny $\Phi^{(1)}$}}=0$.

\section{Final remarks} 
In recent years, the theoretical study of gravitational lenses has been fundamental for the description and analysis of different astrophysical and cosmological phenomena of our Universe. In this work, we have shown how to express different optical scalars and the deflection angle at second order in terms of curvature scalars. These formulas are general, and allow us to find explicit expressions for the optical quantities once the gauge is fixed. 

As an example of the formalism, we have shown that by expressing the Schwarzschild's solution in two different coordinate systems, one obtains at second order the same final expressions for the (asymptotic) optical scalars. This is not surprising, because the evaluation of such scalars in the asymptotic region must only depend from the well-defined quantities as the total ADM mass of the spacetime and the involved impact parameter of the considered null geodesics~\cite{2003AmJPh}.

For a thin lens situation, we also have shown that even when at first order the lens can be thought as if the whole distribution of matter of the lens is placed in a single plane, when one consider second-order corrections, one must be more careful and preserve quadratic terms that at a first sight could be thought that they do not contribute.

\begin{figure}[H]\label{grafico1}
 \centering
\includegraphics[clip,width=95mm]{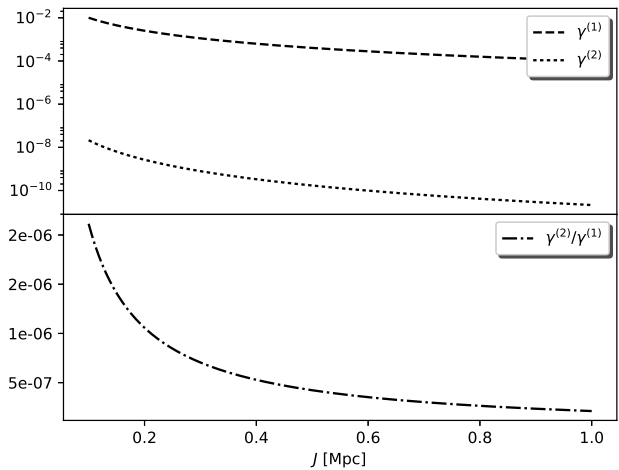}
\caption{Top: a comparative plot of the first and second order contributions to the shear for a spherical mass distribution with $M=10^{12}M_\odot$ situated at a distance of 1500 Mpc and with $\lambda_{s}/\lambda_{l}=2$. Bottom: the quotient between the leading and second order contribution to the shear $\gamma$.}
 \label{grafico1}
 \end{figure}
 
Finally, one would like to make some comments about how important second-order contributions to the optical scalars could be. In order to give an example of the estimated order of magnitude involved, we consider a model of a galaxy as a spherical mass distribution of the order of $10^{12}$ solar masses located at a distance $\lambda_l=1500$Mpc from us. The sources are assumed to be at a distance $\lambda_s=2\lambda_l$.
 Of course, even when a point mass model is a rough model and we  would also take into account cosmological corrections, it nevertheless serves to have an idea of the order of magnitude involved. In fact, compared to more realistic matter distribution models such as those coming from an NFW density profile, it remains a reasonable estimator~\cite{Lasky:2009}. In Figure~\ref{grafico1} we have plotted the contribution that $\gamma^{(2)}$ and $\gamma^{(1)}$ make to the total shear and also its quotient. As it can be seen in the figure, second-order contribution to the shear for this kind of astrophysical system is of the order $10^{-6}$ of the main contribution. A similar crude analysis follows for groups and cluster of galaxies. For example, for the Coma cluster which is placed at 100 Mpc from us, with an estimated  mass of $M_{\text{Coma}}=7\times 10^{14}M_\odot$ at $J\approx 2$ Mpc\cite{Gallo:2011hi}, $\gamma^{(2)}/\gamma^{(1)}\approx 8\times 10^{-5}$.
In general for a point mass, or in the exterior region of a spherical mass the quotient $\gamma^{(2)}/\gamma^{(1)}$ is proportional to the ratio $ {r_{H}/ J}$, with $r_{H}$ the Schwarzschild radius.  

We would also like to mention that in the literature, it can also be found the discussion of gravitational lensing for exotic objects~\cite{Bozza:2015haa, Asada:2017vxl, Kitamura:2013tya, Izumi:2013tya,Tsukamoto:2014dta}. For some of these kind of objects, the first order effect in the optical scalars or deflection angle are of the same order of magnitude that the second order effects coming from a point mass in a Schwarzschild metric. For example, it follows from~\cite{Tsukamoto:2014dta} that for a four-dimensional projection of a five-dimensional Tangherlini spacetime with mass $M$, the deflection angle is proportional to $M^2/J^2$, and therefore the convergence and shear of this metric  in the weak field regime  have a similar functional dependence as the second order contribution to the optical scalars for a four-dimensional Schwarzschild metric with the same mass. 

On the other hand, the contribution of second order weak lensing to the CMB power spectrum or in a general cosmological context is extensively discussed in \cite{Hagstotz:2014qea, 1475-7516-2015-06-050, Kaiser:2015iia,BenDayan:2012wi,Fanizza:2013doa,Marozzi:2014kua, Marozzi:2016uob, Petri:2016qya, DiDio:2014lka,DiDio:2015bua, Schaefer:2005up, Cooray:2002mj,Marozzi:2016qxl,Bernardeau-2010}.
We mention that recently in \cite{Boero:2016nrd},  Boero and Moreschi extended the analysis of Ref.\cite{PhysRevD.83.083007} to the cosmological setting. Due to the relevance of second order effects in the cosmological framework, it would be desirable to generalize some of the results of \cite{Boero:2016nrd} to second order.

\section*{Acknowledgments}
We thank an anonymous referee whose comments and suggestions helped to improve this manuscript.
We acknowledge support from CONICET and SeCyT-UNC.

\appendix
\section{}\label{app:int}
\subsection{Proof of the identity \eqref{identity1}}
In order to prove the identity \eqref{identity1} we only have to integrate by parts the integral in $\lambda{'}$ in the left-hand side taking a function 
\begin{equation}
u=\int_{0}^{\lambda{'}}\lambda{''}Q(\lambda{''})d\lambda{''},
\end{equation}
and the other function $v$ such that 
\begin{equation}
\frac{d v}{d \lambda{'}}=1;
\end{equation}
then, using the well know expression
\begin{equation}\label{intbyparts}
\int u(s) \frac{d v}{ds} ds= u v - \int \frac{d u}{ds} v(s) ds
\end{equation}
and making the identification $\lambda{'}$$\to$$\lambda$ the identity is proven.

\subsection{Proof of the identity \eqref{identity2}}

The identity \eqref{identity2} is a particular case of the following relation:
\begin{equation}\label{identity2a} 
\begin{aligned}
&\int^{\lambda_{s}}_{0} \int^{\lambda{'}}_{0} \int^{\lambda{''}}_{0} \lambda{'''} (\lambda{''}-\lambda{'''})
	f(\lambda{'''},\lambda{''}) d\lambda{'''} d\lambda{''} d\lambda{'} \\
	&= \int_{0}^{\lambda_{s}} \int_{0}^{\lambda{}}\lambda{'}(\lambda_{s}-\lambda{})(\lambda-\lambda{'})f(\lambda{'},\lambda) d\lambda{'}d\lambda{}.
\end{aligned}    
\end{equation}
In order to prove the identity \eqref{identity2a} we have to integrate by parts the integral in $\lambda{'}$ in the left-hand side taking a function
\begin{equation}
u=\int_{0}^{\lambda{'}} \int_{0}^{\lambda{''}}\lambda{'''}(\lambda{''}-\lambda{'''})f(\lambda{'''},\lambda{''}) d\lambda{'''}d\lambda{''},
\end{equation}
and the other function such that 
\begin{equation}
\frac{d v}{d \lambda{'}}=1;
\end{equation}
then, using \eqref{intbyparts} and making the identifications $\lambda{''}$$\to$$\lambda{'}$, $\lambda{'}$$\to$$\lambda$, the identity is proven.
\begin{widetext}
\section{}\label{abcd}
\subsection{Explicit expressions for the coefficients $A$, $B$, $A'$ and $B'$ that appear in the solution of the geodesic deviation equation at second order}

\begin{equation}
\begin{aligned}
A=&  1 - \frac {1}{\lambda_{s}} \int^{\lambda_{s}}_{0} \lambda (\lambda_{s} - \lambda) \Phi_{00}(\lambda) d\lambda 
		 + \frac{1}{\lambda_{s}} \int_{0}^{\lambda_{s}}\int_{0}^{\lambda}\lambda{'}(\lambda_{s}-\lambda)(\lambda-\lambda{'})  
		 \bigg(\Phi_{00}(\lambda)\Phi_{00}(\lambda{'}) + \Psi_{0R}(\lambda) \Psi_{0R}(\lambda{'})  \\
		 &+ \Psi_{0I}(\lambda) \Psi_{0I}(\lambda{'}) \bigg) d\lambda{'} d\lambda 
		 - \frac{1}{\lambda_{s}} \int_{0}^{\lambda_{s}} \lambda (\lambda_{s} - \lambda) \Psi_{0R}(\lambda) d\lambda
		 + \frac{1}{\lambda_{s}} \int_{0}^{\lambda_{s}}\int_{0}^{\lambda}\lambda{'}(\lambda_{s}-\lambda)(\lambda-\lambda{'}) 
		 \bigg(\Phi_{00}(\lambda) \Psi_{0R}(\lambda{'}) \\
		 &+ \Psi_{0R}(\lambda) \Phi_{00}(\lambda{'}) \bigg) d\lambda{'}d\lambda,
\end{aligned}
\end{equation}
\begin{equation}
\begin{aligned}
	B=& - \frac {1}{\lambda_{s}} \int^{\lambda_{s}}_{0} \lambda (\lambda_{s} - \lambda) \Psi_{0I}(\lambda) d\lambda  
	  + \frac{1}{\lambda_{s}} \int_{0}^{\lambda_{s}}\int_{0}^{\lambda}\lambda{'}(\lambda_{s}-\lambda)(\lambda-\lambda{'}) 
	  \bigg( \Phi_{00}(\lambda) \Psi_{0I}(\lambda{'}) + \Psi_{0I}(\lambda) \Phi_{00}(\lambda{'}) \bigg)
	  d\lambda{'}d\lambda \\
	  &- \frac{1}{\lambda_{s}} \int_{0}^{\lambda_{s}}\int_{0}^{\lambda}\lambda{'}(\lambda_{s}-\lambda)(\lambda-\lambda{'}) \bigg( \Psi_{0I}(\lambda) \Psi_{0R}(\lambda{'}) - \Psi_{0R}(\lambda) \Psi_{0I}(\lambda{'}) \bigg) 
	  d\lambda{'}d\lambda,
\end{aligned}
\end{equation}
\begin{equation}
\begin{aligned}
	A{'}=& - \frac {1}{\lambda_{s}} \int^{\lambda_{s}}_{0} \lambda (\lambda_{s} - \lambda) \Psi_{0I}(\lambda) d\lambda  
	  + \frac{1}{\lambda_{s}} \int_{0}^{\lambda_{s}}\int_{0}^{\lambda}\lambda{'}(\lambda_{s}-\lambda)(\lambda-\lambda{'}) 
	  \bigg( \Phi_{00}(\lambda) \Psi_{0I}(\lambda{'}) + \Psi_{0I}(\lambda) \Phi_{00}(\lambda{'}) \bigg) d\lambda{'}d\lambda \\
	  &+ \frac{1}{\lambda_{s}} \int_{0}^{\lambda_{s}}\int_{0}^{\lambda}\lambda{'}(\lambda_{s}-\lambda)(\lambda-\lambda{'}) \bigg( \Psi_{0I}(\lambda) \Psi_{0R}(\lambda{'}) - \Psi_{0R}(\lambda) \Psi_{0I}(\lambda{'}) \bigg) 
	  d\lambda{'}d\lambda,
\end{aligned}
\end{equation}
\begin{equation}
\begin{aligned}
B{'}=& 1 - \frac {1}{\lambda_{s}} \int^{\lambda_{s}}_{0} \lambda (\lambda_{s} - \lambda) \Phi_{00}(\lambda) d\lambda 
		 + \frac{1}{\lambda_{s}} \int_{0}^{\lambda_{s}}\int_{0}^{\lambda}\lambda{'}(\lambda_{s}-\lambda)(\lambda-\lambda{'})  
		 \bigg(\Phi_{00}(\lambda)\Phi_{00}(\lambda{'}) + \Psi_{0R}(\lambda) \Psi_{0R}(\lambda{'})  \\
		 &+ \Psi_{0I}(\lambda) \Psi_{0I}(\lambda{'}) \bigg) d\lambda{'}  d\lambda 
		 + \frac{1}{\lambda_{s}} \int_{0}^{\lambda_{s}} \lambda (\lambda_{s} - \lambda) \Psi_{0R}(\lambda)d\lambda 
		 - \frac{1}{\lambda_{s}} \int_{0}^{\lambda_{s}}\int_{0}^{\lambda}\lambda{'}(\lambda_{s}-\lambda)(\lambda-\lambda{'}) 
		 \bigg(\Phi_{00}(\lambda) \Psi_{0R}(\lambda{'}) \\
		 &+ \Psi_{0R}(\lambda) \Phi_{00}(\lambda{'}) \bigg) d\lambda{'}d\lambda.
\end{aligned}
\end{equation}
\end{widetext}

\section{}\label{A:integral}
\subsection{About the contribution of the quadratic terms in the thin lens approximation}\label{app-thin-lens}

In this Appendix, we show that if we take the naive approximation \eqref{aprox-delgada} to be sufficient in the computation of the quadratic terms, then they should be vanishing.
In (\ref{convergencia:gral2}), (\ref{shear:gral2}), (\ref{rotacion:gral2}) we have terms of the form
\begin{equation}\label{terminos_cuadratios_tipo}
    \int^{\lambda_{s}}_{0} \int^{\lambda}_{0} \lambda{'} (\lambda_{s}-\lambda) (\lambda-\lambda{'}) 
    C(\lambda) D(\lambda{'}) d\lambda{'} d\lambda.
\end{equation}
In order to find the contribution of this expression in the thin lens approximation we proceed as follows. First, we integrate by parts the integral in $\lambda{'}$ taking one function as $u(\lambda{'})=\lambda{'}(\lambda-\lambda{'})$ and the other as $\frac{d v}{d\lambda{'}}=D(\lambda{'})$. The expression \eqref{terminos_cuadratios_tipo} is now reduced to
\begin{equation}\label{a:1}
  \int_{0}^{\lambda_{s}}(\lambda-\lambda_{s})G(\lambda)C(\lambda) d\lambda
\end{equation}
where
\begin{equation}\label{a:1aa}
	G(\lambda) = \int_{0}^{\lambda} (\lambda-2\lambda{'}) \widetilde{D}(\lambda^{'}) d\lambda{'}. 
\end{equation}
Second, we integrate by parts \eqref{a:1} taking $\tilde{u}(\lambda)=(\lambda-\lambda_{s})G(\lambda)$ and $\frac{d \tilde{v}}{d\lambda}=C(\lambda)$:
\begin{equation}\label{rell}
  \begin{aligned}
   &\int^{\lambda_{s}}_{0} \widetilde{C}(\lambda) \bigg[ \int_{0}^{\lambda}(2\lambda{'}+\lambda_{s})\widetilde{D}(\lambda^{'}) d\lambda{'} - \int_{0}^{\lambda}2\lambda\widetilde{D}(\lambda^{'})d\lambda{'} \\
   &- (\lambda_{s}-\lambda)\lambda\widetilde{D}(\lambda)  \bigg] d\lambda
  \end{aligned}
\end{equation}
Finally, implementing the thin lens approximation (\ref{aprox-delgada}) to \eqref{rell} we obtain the anticipated result
\begin{equation}\label{a101}
	\int^{\lambda_{s}}_{0} \int^{\lambda}_{0} \lambda{'} (\lambda_{s}-\lambda) (\lambda-\lambda{'}) 
    C(\lambda) D(\lambda{'}) d\lambda{'} d\lambda = 0.
\end{equation}

\section{}\label{A:sch}
In this Appendix we consider  the second order approximation of the Scharwzschild metric in isotropic and quasi-Minkowskian coordinate systems and also the PPN metric.

We give the expressions of the parallel propagated vectors $\{ \ell^{a},  m^{a}\}$ at first order, the correction to the background geodesics and the required curvature scalars. All these computations were done with the help of the GRTensor package and MAPLE.

These ingredients are necessary to compute the optical scalars and the deflection angle at second order. 

As in the main text, according to \eqref{eq:a.01}, the spherical symmetry allows us to set without loss of generality  
\begin{equation}
	x=J, \ \ \ z=0.
\end{equation}

\subsection{Schwarzschild lens: Isotropic coordinates}
In order to compute the optical scalars and the deflection angle for the second order approximation of the Schwarzschild exterior metric written in isotropic coordinates~\eqref{isotropic} we need to compute the parallel transport of the vectors $\{ \ell^{a},  m^{a} \}$ at first order in $\varepsilon$,
\begin{equation}
\begin{aligned}
 \ell^{t} =& -1 + \bigg( -\frac{1}{\sqrt{J^{2}+(\lambda-\lambda_{l})^{2}}} + \frac{1}{2\sqrt{J^{2}+\lambda_{l}^{2}}} \bigg) \varepsilon\\ &+ \mathcal{O}(\varepsilon^{2}), \\  
 \ell^{x} =& \frac{1}{J} \bigg( \frac{\lambda_{l}-\lambda}{\sqrt{J^{2}+(\lambda-\lambda_{l})^{2}}} 
	  - \frac{\lambda_{l}}{\sqrt{J^{2}+\lambda_{l}^{2}}} \bigg) \varepsilon  + \mathcal{O}(\varepsilon^{2}), \\
 \ell^{y} =& 1 - \frac{\varepsilon}{2\sqrt{J^{2}+\lambda_{l}^{2}}}  
	  + \mathcal{O}(\varepsilon^{2}), \\
 \ell^{z} =& \mathcal{O}(\varepsilon^{2});
\end{aligned} 
\end{equation}
\begin{equation}
\begin{aligned}
m^{t} &= \frac{1}{2\sqrt{2}} \frac{i}{J} \bigg(  \frac{\lambda-\lambda_{l} }{\sqrt{J^{2}+(\lambda-\lambda_{l})^{2}} } 
        + \frac{\lambda_{l} }{\sqrt{J^{2}+\lambda_{l}^{2}}} \bigg) \varepsilon + \mathcal{O}(\varepsilon^{2}), \\
 m^{x} &= \frac{i}{\sqrt{2}} -  \frac{i}{2\sqrt{2}} \frac{\varepsilon}{\sqrt{J^{2}+(\lambda-\lambda_{l})^{2}}}   + \mathcal{O}(\varepsilon^{2}), \\
 m^{y} &= \frac{i}{2\sqrt{2}J} \bigg(  \frac{\lambda-\lambda_{l} }{\sqrt{J^{2}+(\lambda-\lambda_{l})^{2}} } 
        + \frac{\lambda_{l} }{\sqrt{J^{2}+\lambda_{l}^{2}}} \bigg) \varepsilon + \mathcal{O}(\varepsilon^{2}), \\
 m^{z} &= \frac{1}{\sqrt{2}} - \frac{1}{2\sqrt{2}} \frac{\varepsilon}{\sqrt{J^{2}+(\lambda-\lambda_{l})^{2} }}    + \mathcal{O}(\varepsilon^{2}).    
\end{aligned} 
\end{equation}
The correction to the background null geodesic, which follows from the integration of the $\ell^a$ components at first order is
\begin{equation}
\begin{aligned}
 \delta x^{t} =& \bigg( \frac{\lambda}{2\sqrt{J^{2}+\lambda_{l}^{2}}} - \text{arcsinh}(\frac{\lambda-\lambda_{l}}{J}) - \text{arcsinh}(\frac{\lambda_{l}}{J}) \bigg) \varepsilon\\ &+ \mathcal{O}(\varepsilon^{2}), \\  
 \delta x^{x} =& \frac{1}{J} \bigg( \sqrt{J^{2}+\lambda_{l}^{2}}-\sqrt{J^{2}+(\lambda-\lambda_{l})^{2}} 
 - \frac{\lambda\lambda_{l}}{\sqrt{J^{2}+\lambda_{l}^{2}}} \bigg) \varepsilon \\
 &+ \mathcal{O}(\varepsilon^{2}), \\
 \delta x^{y} =& -\frac{1}{2}\frac{\lambda}{\sqrt{J^{2}+\lambda_{l}^{2}}} \varepsilon 
	  + \mathcal{O}(\varepsilon^{2}), \\
 \delta x^{z} =& \mathcal{O}(\varepsilon^{2}).
\end{aligned} 
\end{equation}
The only non-vanishing curvature scalar at second order is $\Psi_{0}$,
\begin{widetext}
\begin{equation}\label{psi-iso-coord}
\begin{aligned}
 \Psi_{0}(J,\lambda) =& \frac{3}{2} \frac{J^2}{(J^{2}+(\lambda - \lambda_{l})^{2})^{5/2}} \varepsilon 
	  - \frac{3}{4}\frac{1}{\sqrt{J^{2}+\lambda_{l}^{2}}(J^{2}+(\lambda-\lambda_{l})^{2})^{7/2}} \bigg[ \sqrt{J^{2}+\lambda_{l}^{2}} \sqrt{J^{2}+(\lambda-\lambda_{l})^{2}} \\
	  &\times ( J^{2}-4\lambda^{2} + 8\lambda\lambda_{l} -4\lambda_{l}^{2} ) + 2J^{2} ( J^{2}+3\lambda_{l}^{2} + \lambda^{2} -4\lambda\lambda_{l} ) + 4\lambda_{l} (3\lambda^{2}\lambda_{l} - 3\lambda\lambda_{l}^{2}+ \lambda_{l}^{3} - \lambda^{3})
	   \bigg] \varepsilon^{2} 
	  + \mathcal{O}(\varepsilon^{3}).
\end{aligned}
\end{equation}
\end{widetext}
\begin{widetext}
\subsection{Schwarzschild lens: Quasi-Minkoskian coordinates}\label{apendiceA}
In order to compute the optical scalars and the deflection angle for the Schwarzschild point mass lens written in quasi-Minkowskian coordinates~\eqref{cartesian-coord} we need to compute the parallel transport of the vectors $\{ \ell^{a},  m^{a} \}$ at first order in $\varepsilon$:
\begin{equation}
\begin{aligned}
 \ell^{t} =& -1 + \bigg( -\frac{1}{\sqrt{J^{2}+(\lambda-\lambda_{l})^{2}}} + \frac{1}{\sqrt{J^{2}+\lambda_{l}^{2}}} \bigg) \varepsilon \\&+ \mathcal{O}(\varepsilon^{2}),\\  
 \ell^{x} =& \frac{1}{2J} \bigg( \frac{3 J^{2}(\lambda_{l}-\lambda) + 2 (\lambda_{l}-\lambda)^{3}}{(J^{2}+(\lambda-\lambda_{l})^{2})^{3/2}}
	  - \frac{3\lambda_{l} J^{2} + 2 \lambda_{l}^{3}}{(J^{2}+\lambda_{l}^{2})^{3/2}} \bigg) \varepsilon\\ &+ \mathcal{O}(\varepsilon^{2}), \\
 \ell^{y} =& 1 + \frac{ J^{2}}{2} \bigg( \frac{1}{(J^{2}+(\lambda-\lambda_{l})^{2})^{3/2}} - \frac{1}{(J^{2}+\lambda_{l}^{2})^{3/2}} \bigg) \varepsilon\\ 
	  &+ \mathcal{O}(\varepsilon^{2}), \\
\ell^{z} &= \mathcal{O}(\varepsilon^{2}); \\
\end{aligned} 
\end{equation}
and
\begin{equation}
\begin{aligned}
 m^{t} =& \frac{1}{2\sqrt{2}} \frac{i}{J} \bigg( - \frac{(\lambda_{l}- \lambda) }{\sqrt{J^{2}+(\lambda-\lambda_{l})^{2}} } 
        + \frac{\lambda_{l} }{\sqrt{J^{2}+\lambda_{l}^{2}}} \bigg) \varepsilon\\ &+ \mathcal{O}(\varepsilon^{2}), \\
  m^{x} =& \frac{i}{\sqrt{2}} +  \frac{i \, J^{2} }{2\sqrt{2}} \bigg( -\frac{1}{(J^{2}+(\lambda-\lambda_{l})^{2})^{3/2}} \\
  &+  \frac{1}{(J^{2}+\lambda_{l}^{2})^{3/2}} \bigg) \varepsilon+ \mathcal{O}(\varepsilon^{2}), \\
  m^{y} =& \frac{1}{2\sqrt{2}} \frac{i}{J} \bigg( \frac{(\lambda - \lambda_{l})^{3} }{(J^{2}+(\lambda-\lambda_{l})^{2})^{3/2} } 
         + \frac{\lambda_{l}^{3} }{(J^{2}+\lambda_{l}^{2})^{3/2}} \bigg) \varepsilon\\ &+ \mathcal{O}(\varepsilon^{2}), \\
 m^{z} =& \frac{i}{\sqrt{2}} + \mathcal{O}(\varepsilon^{2}). \\    
\end{aligned} 
\end{equation}
The first order contribution to the actual path is
\begin{equation}
\begin{aligned}
 \delta x^{t} =& \bigg( \frac{\lambda}{\sqrt{J^{2}+\lambda_{l}^{2}}} - \text{arcsinh}(\frac{\lambda-\lambda_{l}}{J}) - \text{arcsinh}(\frac{\lambda_{l}}{J}) \bigg) \varepsilon+ \mathcal{O}(\varepsilon^{2}), \\  
 \delta x^{x} =& \frac{1}{2J\sqrt{J^{2}+(\lambda-\lambda_{l})^{2}}(J^{2}+\lambda_{l}^{2})^{3/2}} \bigg(\sqrt{J^{2}+\lambda_{l}^{2}}
 (4J^{2}\lambda\lambda_{l}-J^{4}-3J^{2}\lambda_{l}^{2}-2J^{2}\lambda^{2}
 -2\lambda^{2}\lambda_{l}^{2}
 +4\lambda\lambda_{l}^{3}-2\lambda_{l}^{4})\\
 &+\sqrt{J^{2}+(\lambda-\lambda_{l})^{2}} 
 (J^{4}
 +3J^{2}\lambda_{l}^{2}-3J^{2}\lambda\lambda_{l}+2\lambda_{l}^{4}-2\lambda\lambda_{l}^{3}) \bigg) \varepsilon + \mathcal{O}(\varepsilon^{2}), \\
 \delta x^{y} =& \frac{1}{2\sqrt{J^{2}+(\lambda-\lambda_{l})^{2}}(J^{2}+\lambda_{l}^{2})^{3/2}} \bigg(\sqrt{J^{2}+\lambda_{l}^{2}} (J^{2}\lambda+\lambda\lambda_{l}^{2}-J^{2}\lambda_{l}-\lambda_{l}^{3})+\sqrt{J^{2}+(\lambda-\lambda_{l})^{2}}\bigg)(J^{2}\lambda_{l}+\lambda_{l}^{3}-J^{2}\lambda) \varepsilon \\
 &+ \mathcal{O}(\varepsilon^{2}), \\
 \delta x^{z} =& \mathcal{O}(\varepsilon^{2}).
\end{aligned} 
\end{equation}
While for the curvature quantity \eqref{psipsi} we get
\begin{equation}
\begin{aligned}
 \Psi_{0}(J,\lambda) =& \frac{3}{2} \frac{J^{2}}{(J^{2}+(\lambda - \lambda_{l})^{2})^{5/2}} \varepsilon  
	  + \frac{3}{2} \frac{1}{\sqrt{J^{2}+\lambda_{l}^{2}}(J^{2}+(\lambda - \lambda_{l})^{2})^{7/2}} 
	   \bigg( 2 (\lambda - \lambda_{l})^{2} \sqrt{J^{2}+\lambda_{l}^{2}} \sqrt{J^{2}+(\lambda - \lambda_{l})^{2}} \\
	  &+ J^{2} \sqrt{J^{2}+\lambda_{l}^{2}} \sqrt{J^{2}+(\lambda - \lambda_{l})^{2}} 
	  + 4 \lambda_{l} \lambda J^{2} - 3 \lambda_{l}^{2} J^{2} + 2 \lambda_{l} (\lambda - \lambda_{l})^{3} 
	  - J^{4} - \lambda^{2} J^{2} \bigg) \varepsilon^{2} + \mathcal{O}(\varepsilon^{3}).
\end{aligned}
\end{equation}
Despite the above expression for $\Psi_{0}$ is different of \eqref{psi-iso-coord} which is calculated in isotropic coordinates, the expressions for the convergence, shear and deflection angle are the same in the limits $\lambda_{l} \to \infty $, $ \lambda_{ls} \to \infty$.
\end{widetext}
\begin{widetext}
\subsection{Parametrized-post-Newtonian point mass lens}

For the PPN point mass metric the components of $\{\ell^{a},m^{a}\}$ and the contribution to the actual geodesic $\delta x^{a}$ at first order are given by:
\begin{equation}
\begin{aligned}
 \ell^{t} =& -1 + \bigg( -\frac{1}{\sqrt{J^{2}+(\lambda-\lambda_{l})^{2}}} + \frac{1}{2\sqrt{J^{2}+\lambda_{l}^{2}}} \bigg) \varepsilon\\ &+ \mathcal{O}(\varepsilon^{2}), \\  
 \ell^{x} =& \frac{1}{2} \frac{(\mu + 1)}{J} \bigg( \frac{\lambda_{l}-\lambda}{\sqrt{J^{2}+(\lambda-\lambda_{l})^{2}}} 
	  - \frac{\lambda_{l}}{\sqrt{J^{2}+\lambda_{l}^{2}}} \bigg) \varepsilon\\  &+ \mathcal{O}(\varepsilon^{2}), \\
 \ell^{y} =& 1 - \frac{1}{2}\bigg( \frac{(\mu-1)}{\sqrt{J^{2}+(\lambda-\lambda_{l})^{2}}} + \frac{1}{\sqrt{J^{2}+\lambda_{l}^{2}}} \bigg) \varepsilon 
	  + \mathcal{O}(\varepsilon^{2}), \\
 \ell^{z} =& \mathcal{O}(\varepsilon^{2});
\end{aligned} 
\end{equation}
\begin{equation}
\begin{aligned}
m^{t} &= \frac{1}{2\sqrt{2}} \frac{i}{J} \bigg(  \frac{\lambda-\lambda_{l} }{\sqrt{J^{2}+(\lambda-\lambda_{l})^{2}} } 
        + \frac{\lambda_{l} }{\sqrt{J^{2}+\lambda_{l}^{2}}} \bigg) \varepsilon + \mathcal{O}(\varepsilon^{2}), \\
 m^{x} &= \frac{i}{\sqrt{2}} -  \frac{i \, \mu}{2\sqrt{2}} \frac{\varepsilon}{\sqrt{J^{2}+(\lambda-\lambda_{l})^{2}}}   + \mathcal{O}(\varepsilon^{2}), \\
 m^{y} &= \frac{\mu}{2\sqrt{2}} \frac{i}{ J} \bigg(  \frac{\lambda-\lambda_{l} }{\sqrt{J^{2}+(\lambda-\lambda_{l})^{2}} } 
        + \frac{\lambda_{l} }{\sqrt{J^{2}+\lambda_{l}^{2}}} \bigg) \varepsilon + \mathcal{O}(\varepsilon^{2}), \\
 m^{z} &= \frac{1}{\sqrt{2}} - \frac{\mu}{2\sqrt{2}} \frac{\varepsilon}{\sqrt{J^{2}+(\lambda-\lambda_{l})^{2} }}    + \mathcal{O}(\varepsilon^{2});     
\end{aligned} 
\end{equation}
\begin{equation}
\begin{aligned}
 \delta x^{t} =& \bigg( \frac{\lambda}{2\sqrt{J^{2}+\lambda_{l}^{2}}} - \text{arcsinh}(\frac{\lambda-\lambda_{l}}{J}) - \text{arcsinh}(\frac{\lambda_{l}}{J}) \bigg) \varepsilon\\ &+ \mathcal{O}(\varepsilon^{2}), \\  
 \delta x^{x} =& \frac{1}{2} \frac{(\mu + 1)}{J} \bigg( \sqrt{J^{2}+\lambda_{l}^{2}}-\sqrt{J^{2}+(\lambda-\lambda_{l})^{2}} \\
 &- \frac{\lambda\lambda_{l}}{\sqrt{J^{2}+\lambda_{l}^{2}}} \bigg) \varepsilon + \mathcal{O}(\varepsilon^{2}), \\
 \delta x^{y} =& \frac{1}{2}\bigg[ (1-\mu) \bigg(\text{arcsinh}(\frac{\lambda-\lambda_{l}}{J}) + \text{arcsinh}(\frac{\lambda_{l}}{J}) \\
 &- \frac{\lambda}{\sqrt{J^{2}+\lambda_{l}^{2}}} \bigg) - \frac{\mu\lambda}{\sqrt{J^{2}+\lambda_{l}^{2}}}\bigg] \varepsilon 
	  + \mathcal{O}(\varepsilon^{2}), \\
 \delta x^{z} =& \mathcal{O}(\varepsilon^{2});
\end{aligned} 
\end{equation}
and the curvature quantities $\{\Psi_{0},\Phi_{00}\}$ at second order are given by:
\begin{equation}
\begin{aligned}
\Psi_{0}(J,\lambda) =& \frac{3}{4} \frac{(\mu + 1) J^2}{(J^{2}+(\lambda - \lambda_{l})^{2})^{5/2}} \varepsilon 
	  + \frac{1}{8}\frac{1}{\sqrt{J^{2}+\lambda_{l}^{2}}(J^{2}+(\lambda-\lambda_{l})^{2})^{7/2}} \bigg[ \sqrt{J^{2}+\lambda_{l}^{2}} \sqrt{J^{2}+(\lambda-\lambda_{l})^{2}} \\
	  &\times \bigg( 6 (\lambda - \lambda_{l})^{2} + J^{2} (13 - 8\beta - 2\mu -15\mu^{2} + 6\nu) + 6\mu ( \mu \lambda_{l}^{2} + 2 \lambda^{2} 
	  + \lambda^{2} \mu + 2 \lambda_{l}^{2} - 2 \lambda \lambda_{l} \mu - 4 \lambda \lambda_{l}) \bigg) \\
	  &- 6 J^{2} ( \lambda_{l}^{2}\mu^{2}-\lambda \lambda_{l}\mu^{2}-4\lambda_{l}\lambda\mu+\mu\lambda^{2}+3\lambda_{l}^{2}\mu-3\lambda_{l}\lambda+\lambda^{2} ) + 6\mu^{2} ( \lambda_{l}\lambda^{3}+3\lambda_{l}^{3}\lambda-3\lambda_{l}^{2}\lambda^{2}-\lambda_{l}^{4} ) \\
      &+ 12\mu ( 3 \lambda \lambda_{l}^{3} - 3 \lambda^{2}\lambda_{l}^{2}  + \lambda^{3} \lambda_{l} - \lambda_{l}^{4} -\frac{1}{2}J^{4}) 
	  - 6 ( \lambda_{l}^{4} + J^{4} -3\lambda_{l}^{3}\lambda+3\lambda_{l}^{2}\lambda^{2}-\lambda_{l}\lambda^{3} ) \bigg] \varepsilon^{2} 
	  + \mathcal{O}(\varepsilon^{3});
\end{aligned}
\end{equation}
\begin{equation}
\begin{aligned}
 \Phi_{00}(J,\lambda) =& \frac{1}{4} \frac{(\mu - 1) (J^{2} - 2 (\lambda - \lambda_{l})^{2})}{(J^{2}+(\lambda- \lambda_{l})^{2})^{5/2}} \varepsilon 
			+ \frac{1}{8} \frac{1}{\sqrt{J^{2}+\lambda_{l}^{2}} (J^{2}+(\lambda- \lambda_{l})^{2})^{7/2}} \bigg[ \sqrt{J^{2} + \lambda_{l}^{2}} 
			\sqrt{J^{2} + (\lambda - \lambda_{l})^{2}} \\
			&\times \bigg( 2(1 - 2\beta - 3\nu) (\lambda-\lambda_{l})^{2} + J^{2} (4\mu - 3\mu^{2} + 4\beta - 5) + 2\mu (9\mu \lambda_{l}^{2} 
			- 18\mu \lambda \lambda_{l} + 9\mu \lambda^{2} - 5 \lambda_{l}^{2} + 10 \lambda \lambda_{l} \\
            &- 5 \lambda^{2} ) \bigg) 
			+ 2 J^{2} ( \lambda_{l}^{2}\mu- \lambda_{l}\lambda + 3\mu^{2}\lambda_{l}\lambda + \mu\lambda^{2} - \lambda^{2} + 2\lambda_{l}^{2} - 2\mu\lambda_{l}\lambda -      			     3\mu^{2}\lambda_{l}^{2} ) - 6\mu^{2} ( 3\lambda_{l}^{2}\lambda^{2}-3\lambda_{l}^{3}\lambda - \lambda_{l} \lambda^{3} \\
            &+ \lambda_{l}^{4} ) 
			- 4\mu ( 4\lambda_{l}\lambda^{3} + 4\lambda_{l}^{3}\lambda - 6\lambda_{l}^{2}\lambda^{2} + \frac{1}{2}J^{4}- \lambda_{l}^{4} - \lambda^{4} )
			- 2 ( 3\lambda_{l}^{2}\lambda^{2} - J^{4} - \lambda_{l}^{4} + \lambda_{l}^{3}\lambda + 2\lambda^{4} - 5\lambda_{l}\lambda^{3} ) \bigg] \varepsilon^{2} \\
            &+ \mathcal{O}(\varepsilon^{3}).
\end{aligned}
\end{equation}
\end{widetext}

\bibliographystyle{unsrt}

\end{document}